\documentclass{aa}

\usepackage{graphicx
}

\usepackage{booktabs}

\usepackage{txfonts}

\newcommand{\Msun}{\rm{M}_{\odot}}

\begin{document}

   \title{Primordial black holes as dark matter candidates}

   \subtitle{Multi-frequency constraints from cosmic radiation backgrounds}

   \author{C. Casanueva-Villarreal\inst{1,2}
          \and
          N. Padilla\inst{3,4}
          \and
          P. B. Tissera\inst{1,2}
          \and
          B. Liu\inst{5}
          \and
          V. Bromm\inst{6}
          }

   \institute{Instituto de Astrofísica, Pontificia Universidad Católica de Chile, Av. Vicuña Mackenna 4860, Santiago, Chile\
         \and
            Centro de Astro-Ingeniería, Pontificia Universidad Católica de Chile, Av. Vicuña Mackenna 4860, Santiago, Chile
        \and
            Instituto de Astronomía Teórica y Experimental (IATE), CONICET-Universidad Nacional de Córdoba, Laprida 854, X5000BGR, Córdoba, Argentina
        \and 
            Observatorio Astronómico de la Universidad Nacional de Córdoba, Laprida 854, X5000BGR, Córdoba, Argentina
        \and
            Institut für Theoretische Astrophysik, Zentrum für Astronomie, Universität Heidelberg, D-69120 Heidelberg, Germany
        \and
            Department of Astronomy, The University of Texas at Austin, Austin, TX 78712, USA}

   \date{Received September 15, 1996; accepted March 16, 1997}

  \abstract
   {}
   {This study investigates the role of primordial black holes (PBHs) in shaping cosmic radiation backgrounds, specifically the cosmic X-ray background (CXB), the Lyman-Werner background (LWB), and the cosmic radio background (CRB). It assesses their viability as dark matter (DM) candidates based on both observational constraints and theoretical limits.
}
   {PBH accretion is modelled using analytical frameworks, including electron advection-dominated accretion flows (eADAF), standard ADAF, luminous hot accretion flows (LHAF), and thin disks. Contributions to the CXB, LWB, and CRB are calculated for PBHs in both halos and the intergalactic medium (IGM). To test robustness, we explore variations in the model, such as halo density profiles, gas velocities and emission models. The results are compared against observational limits and theoretical thresholds across these backgrounds, constraining the PBH fraction as DM for masses between 1 and 100 M$_{\odot}$. 
}
   {Our findings suggest that PBHs can contribute up to 99, 93, 80, and 91 per cent of the observed non-source soft X-ray background for masses of \(1 \, \Msun\), \(10 \, \Msun\), \(33 \, \Msun\), and \(100 \, \Msun\), respectively, while contributing approximately 33, 37, 33, and 39 per cent to the hard X-ray background. These contributions constrain the maximum DM fraction in the form of PBHs to \(7 \times 10^{-3}\), \(6 \times 10^{-4}\), \(6 \times 10^{-4}\), and \(7 \times 10^{-4}\) for the respective masses under the baseline model. These constraints align with the limits imposed by the LWB, ensuring that PBHs do not disrupt molecular cooling or early star formation under these conditions. However, explaining the observed radio background excess at \(z = 0\) and the EDGES signal would require DM fractions composed of PBHs significantly larger than those allowed by these constraints. For \(1 \, \Msun\), excluding subregimes in the ADAF framework relaxes the constraint to \(3 \times 10^{-2}\), highlighting the impact of the modelled accretion physics on the derived limits.
Variations in model assumptions, such as halo density profiles, gas velocities, emission models, and modifications to the halo mass function, introduce slight changes in the predicted backgrounds. 
}
   {}


   \maketitle

\section{Introduction}

Cosmic radiation backgrounds serve as integral probes of the energetic and dynamic processes that shaped the early Universe. These diffuse emissions, spanning from radio to high-energy X-rays, encapsulate the integrated contributions of astrophysical sources over cosmic time. Notably, the cosmic X-ray background (CXB), cosmic radio background (CRB), and the Lyman-Werner background (LWB) provide critical insights into energetic astrophysical processes, the interstellar medium (ISM), and the regulation of star formation in the early Universe. Observations have reported significant excesses in the CXB and CRB beyond the expected contributions from known astrophysical sources \citep{Fixsen_2011,Cappelluti_2017}. At the same time, the LWB plays a crucial role in regulating star formation by dissociating molecular hydrogen (H$_2$), the primary coolant in metal-poor gas during the early universe \citep{Stecher_Williams_1967,Saslaw_Zipoy_1967,Peebles_Dicke_1968}.

The CXB, primarily measured by satellite missions such as the Chandra X-ray Observatory (Chandra), the Röntgen Satellite (ROSAT), and the X-ray Multi-Mirror Mission (XMM-Newton), traces high-energy processes associated with active galactic nuclei (AGNs), star-forming galaxies, and high-redshift black holes \citep{Gilli_2007}. Although most of the CXB can be attributed to known populations, observations reveal an unresolved excess in the $0.5$–$10$ keV band. This excess is quantified as non-source contributions (nsCXB), defined as the component of the CXB that remains after masking all sources detected in X-ray surveys and those with counterparts identified in other wavelengths. nsCXB accounts for $9.7^{+1.6}_{-1.8}$ per cent of the total CXB in the $0.5$–$2$ keV range and $17^{+5.9}_{-7.0}$ per cent in the $2$–$10$ keV range \citep{Cappelluti_2017}\footnote{Recent JWST results suggest that a substantial fraction of the unresolved CXB (uCXB)—the component remaining after subtracting detected X-ray sources—can be attributed to high-redshift galaxies \citep{Kaminsky_2025}. However, our analysis relies on the non-source CXB (nsCXB), defined as the residual after masking all sources detected in X-rays and in multiwavelength data \citep{Cappelluti_2017}. Since the nsCXB is a subset of the uCXB, the newly resolved sources may reduce its inferred level, but this depends on whether they overlap with those contributing to the nsCXB. Further analysis is required to assess how JWST impacts nsCXB estimates and, consequently, the derived PBH constraints. We defer this investigation to future work. Importantly, even with recent identifications, some room remains for an unresolved excess.}. Potential contributors include heavily obscured, Compton-thick AGNs, which still avoid detection with traditional X-ray surveys, and star-forming galaxies at high redshifts. However, their contribution is estimated to be minor. Population synthesis models, which combine AGN luminosity functions and spectral properties, suggest that these sources can only partially explain the nsCXB, leaving a significant fraction unresolved \citep{Ananna_2020}. Ongoing and upcoming missions, such as the Nuclear Spectroscopic Telescope Array (NuSTAR; \citealt{Harrison_2013}) and the High-Energy X-ray Probe (HEX-P; \citealt{Civano_2024}), are expected to provide sensitive broad-band X-ray observations, helping to refine these models and uncover the nature of the remaining unresolved sources.

The CRB observed using instruments such as the Absolute Radiometer for Cosmology, Astrophysics, and Diffuse Emission (ARCADE 2) is primarily composed of contributions from well-known astrophysical sources, including the cosmic microwave background (CMB), radio-loud AGNs, and star-forming galaxies. However, ARCADE 2 measurements reveal a puzzling excess at frequencies between $3-8$ GHz, with a reported temperature rise of $54 \pm 6$ mK at 3.3 GHz \citep{Fixsen_2011}. This excess remains unexplained by known sources, leading to exploring alternative explanations. Proposed hypotheses include synchrotron radiation from highly efficient, accreting supermassive black holes (SMBHs; \citealt{Ewall-Wice_2018}) and extreme populations of supernova remnants \citep{Mirocha_Furlanetto_2019}. More exotic processes have also been suggested, such as synchrotron radiation emitted by relativistic electrons through late decays of a metastable particle \citep{Cline_2013}, decays of dark matter (DM) particles resulting in dark photons oscillating into ordinary photons \citep{Caputo_2023}, and radiative decays of relic neutrinos into sterile neutrinos \citep{Bhupal_2024}. However, these scenarios require extreme efficiencies or unconventional physics that remain challenging to reconcile with current observations. Further investigations into the origins of the CRB excess are necessary to fully understand its implications for cosmology and the physics of diffuse backgrounds.

The LWB, consisting of ultraviolet photons with energies between 11.2 eV and 13.6 eV, plays a pivotal role in shaping the early Universe by regulating star formation. These photons, mainly emitted by massive stars, dissociate molecular hydrogen (H$_2$) through the Solomon process \citep{Stecher_Williams_1967}, suppressing the primary cooling mechanism in metal-poor gas \citep{Saslaw_Zipoy_1967, Peebles_Dicke_1968}. Molecular cooling becomes inefficient above temperatures of $\sim10^4$ K, where atomic cooling processes dominate, particularly in halos with sufficient gravitational potential to sustain such mechanisms \citep{Haiman_Thoul_1996, Machacek_2001}. The intensity of the LWB and its interaction with self-shielding effects in dense regions determine the extent of this suppression \citetext{e.g., \citealt{Johnson_2008}}. In regions where self-shielding is effective, H$_2$ molecules can survive, allowing localised cooling and star formation to proceed \citep{Wolcott-Green_2011, Hartwig_2015}. This typically occurs in halos with masses above $\sim 10^6 - 10^7 \, M_\odot$, where higher gas densities enable efficient self-shielding against the dissociating LW radiation. However, in diffuse environments, such as lower-mass halos below this threshold, the LWB efficiently dissociates molecular hydrogen, delaying or suppressing star formation. This feedback mechanism increases the critical halo mass threshold for star formation, governing the chemothermal evolution of the primordial gas \citep{Latif_2019, Schauer_2021}. Beyond its role in star formation, the LWB sets the conditions necessary for the direct collapse of gas into black hole seeds, which may serve as precursors to SMBHs observed at high redshifts \citep{Agarwal_2019, Wise_2019}.

Primordial black holes (PBHs) have emerged as a compelling hypothesis to explain the observed anomalies in the CXB and CRB, as well as potential contributors to the LWB. These objects, theorised to form from density fluctuations in the early Universe, have been proposed as candidates for DM and could constitute a significant fraction of its total density \citep{Carr_2016}. Unlike stellar black holes, PBHs are not limited by the processes of stellar collapse and can span a wide range of masses, offering unique insights into early universe physics. Their accretion processes produce radiation across the electromagnetic spectrum, including the X-ray, radio, and ultraviolet bands \citep{Ricotti_2008, Mack_2007}. Specifically, accreting PBHs could generate X-rays and synchrotron emission from relativistic electrons, partially accounting for the observed excesses in the CXB and CRB \citep{Hasinger_2020,Cappelluti_2022}. In addition, PBHs could influence the LWB by emitting ultraviolet radiation during gas accretion, further regulating molecular hydrogen dissociation and impacting star formation in low-mass halos.

Recent James Webb Space Telescope (JWST) discoveries have significantly expanded our understanding of the early Universe, revealing numerous massive galaxies and supermassive black holes (SMBHs) at high redshifts. Many galaxy candidates at $z \gtrsim 10$ have been identified using NIRCam imaging through programs such as Early Release Observations (ERO; \citealt{Pontoppidan_2022}) and Director’s Discretionary Early Release Science (ERS). Examples include GN-z11, identified at $z=10.6$ \citep{Tacchella_2023}, which exhibits a higher-than-expected stellar mass. Such discoveries, if correct, challenge the standard $\Lambda$CDM cosmology by highlighting the rapid formation of massive structures in the early universe (e.g., \citealt{Atek_2022, Castellano_2022, Finkelstein_2022, Naidu_2022, Yan_2022, Bradley_2023, Robertson_2023, Yan_2023}). Similarly, SMBHs with masses exceeding $10^7 \Msun$ have been observed at $z=7-10$, suggesting growth rates that are difficult to reconcile with traditional black hole formation models \citep{Goulding_2023, Larson_2023, Bogdan_2024, Greene_2024, Natarajan_2024}. These findings suggest that PBHs could act as seeds for SMBHs, accelerating their growth and contributing to the rapid formation of massive galaxies in the early universe \citep{CarrKuhnel_2020}. An extreme example of such a BH-dominated high-redshift system discovered by JWST is Abell 2744-QSO1, which is overmassive even with respect to the host halo dynamical mass (as opposed to its stellar mass), challenging standard formation channels \citep{Ji_2025}.

Simulations further support this hypothesis. \cite{Colazo24} demonstrated that even a small abundance of PBHs, accounting for only 0.5 per cent of the DM, with extended mass functions from \cite{Sureda_2021}, can provide the seeds for SMBHs and enhance the abundance of high-redshift galaxies observed by JWST. Their study, employing the SWIFT simulation code, highlights how PBHs can drive structure formation in the early Universe with realistic star formation efficiencies. Additionally, \cite{Liu_2022b} showed, through an analytical model based on linear perturbation theory and the Press-Schechter (PS) formalism, that massive PBHs (\( \gtrsim 10^9 ~\mathrm{M}_\odot \)), assuming a monochromatic mass function and comprising \( \sim 10^{-6} - 10^{-3} \) of the DM, can significantly accelerate the formation of massive galaxies at \( z \gtrsim 10 \). Their analysis demonstrated that the gravitational potential of such PBHs enhances structure formation by reducing the star formation efficiency required to explain the observed abundance of massive galaxies at high redshifts, aligning theoretical predictions with JWST observations. In addition, \cite{Liu_2022}, using cosmological hydrodynamic zoom-in simulations and semi-analytical models, investigated the impact of stellar-mass PBHs on the first generation of stars, showing that PBHs amplify small-scale density perturbations and influence early star formation through accretion feedback. These models not only align with observational constraints on PBH parameters but also predict potential signatures that JWST could confirm in future studies, offering a promising avenue for testing the role of PBHs in cosmic evolution. Collectively, these findings underscore the ability of PBHs, even in minimal fractions, to serve as seeds for SMBHs and play a pivotal role in shaping the formation and evolution of massive structures in the early Universe. While PBHs are a compelling candidate to explain the observed abundance of massive galaxies and SMBHs at high redshifts, other hypotheses have also been proposed. For instance, \cite{Koehler_2024} explored the role of cosmic string loops in seeding galaxy formation, demonstrating through cosmological simulations that they could also contribute to explaining JWST observations.

Focusing on cosmic radiation backgrounds, \cite{Ziparo_2022} investigated emissions produced by PBHs, particularly in the CXB and CRB. Using the Bondi–Hoyle–Lyttelton accretion model, they calculated the accretion rate based on gas density, sound speed, and the relative velocity between the PBH and the surrounding gas. Their findings indicate that PBHs can produce significant emissions in X-ray and radio bands through accretion. However, their results suggest that while PBHs could account for a fraction of the CXB, they cannot fully explain the CRB excess observed by ARCADE 2.

Similarly, PBHs have been explored as a possible explanation for the anomalous 21 cm absorption signal reported by the Experiment to Detect the Global Epoch of Reionisation Signature \citep[EDGES;][]{Bowman_2018, Ewall-Wice_2018}. However, this detection remains debated, and it is important to acknowledge that the Shaped Antenna Measurement of the Background Radio Spectrum (SARAS) experiment found no evidence supporting the signal and suggested that systematic effects might be responsible for the observed feature \citep{Singh_2022}. Despite this controversy, the EDGES signal is not definitively ruled out and, therefore, continues to be investigated in theoretical studies, particularly in relation to PBH-driven heating. In this context, \cite{Ziparo_2022} showed that PBHs could impact the 21 cm signal by heating the intergalactic medium (IGM) through X-ray emission and by contributing to a potential excess radio background; however, their results indicate that the conditions required for PBHs to reproduce the observed EDGES signal are tightly constrained by existing observational limits.

Our previous work, \cite{Casanueva_2024}, hereafter CV2024, investigated the role of PBHs in the early Universe by analysing their impact on gas properties at very high redshifts ($z \sim 23$). This study employed a combination of hydrodynamical simulations, semi-analytical models, and analytical approaches to explore various accretion regimes, including standard advection-dominated accretion flow (ADAF), electron ADAF (eADAF), luminous hot accretion flow (LHAF), and thin disc models. This comprehensive analysis revealed how the temperature and hydrogen abundance near PBH-hosting halos depend on different DM scenarios, offering a more nuanced understanding of PBH accretion processes and their influence on early galaxy formation.

Despite their theoretical potential, PBHs remain heavily constrained by multiple observations. For example, \cite{Ziparo_2022} concluded that while PBHs could contribute to the CXB, their model could not fully account for the observed CRB. In addition, various multi-messenger observations have placed stringent bounds on the allowed PBH abundance. These include microlensing constraints from surveys such as Subaru \citep{Niikura_2019b, Kusenko_2020}, EROS-2/MACHO \citep{Tisserand_2007}, and OGLE \citep{Niikura_2019a, Mroz_2024}, gravitational wave limits from LIGO/Virgo \citep{Abbott_2019, Wong_2021}, constraints from CMB spectral distortions due to accretion \citep{Poulin_2017}, and limits from radio and X-ray observations of the Galactic Centre \citep{Manshanden_2019}. Together, these independent constraints span a broad mass range and significantly restrict the parameter space in which PBHs could constitute a substantial fraction of DM (see \citealt{Carr+2020} for a review). Additionally, while PBHs might play a role in SMBH formation and galaxy assembly, further observational and theoretical refinement is needed to evaluate their overall impact comprehensively; for a detailed discussion, see \cite{LiuBromm2023}.

This study aims to constrain the fraction of DM that can be composed of PBHs by examining the radiation backgrounds they produce. Specifically, we compare the results considering different gas density profiles within halos to understand how these profiles affect the predicted radiation and the resulting constraints on the PBH contribution to DM. Building on CV2024, we analyse PBH accretion across various sub-regimes, such as eADAF, ADAF, LHAF, and thin disc models. This approach not only refines constraints on PBH parameters but also sheds light on their potential role as DM candidates and their broader influence on early universe cosmology.

This paper is structured as follows. Section \ref{sec:baseline_model} outlines the theoretical framework of this study. In Section \ref{subsubsec:distribution}, we describe the cosmological distribution of PBHs, including their interplay with the large-scale structure of the Universe. Section \ref{subsec:accretion} focuses on modelling accretion processes, highlighting the different regimes and their implications for radiation production. Section \ref{sec:results} presents the main results, evaluating the contributions of PBHs to the X-ray, Lyman-Werner, and radio backgrounds and comparing these to current observational constraints. In Section \ref{subsec:model_variations}, we explore the impact of varying key modelling assumptions on our results to assess their robustness. Finally, Section \ref{sec:conclusion} synthesises the findings and outlines prospects for future research.

\section{Model}\label{sec:baseline_model}

In this section, we elaborate on the theoretical framework employed in our study to examine the effects of PBHs on cosmic radiation backgrounds. Our model incorporates the cosmological distribution of PBHs as well as the physical processes associated with their accretion and the resulting radiation.

We adopt a $\Lambda$CDM cosmology consistent with the Planck 2018 results \citep{Planck_2020}: \( \Omega_{\rm{m}} = 0.315 \), \( \Omega_\Lambda = 0.685 \), \( \Omega_{\rm{b}} = 0.049 \), \( \sigma_8 = 0.811 \), \( n_{\rm{s}} = 0.965 \), and \( H_0 = 67.4 \, \text{km} \, \text{s}^{-1} \, \text{Mpc}^{-1} \).

\subsection{Cosmological distribution and structure formation}
\subsubsection{Cosmological distribution of PBHs}\label{subsubsec:distribution}

We assume a monochromatic mass distribution of PBHs which constitute a fraction $f_{\text{PBH}}$ of DM, defined as \( f_{\text{PBH}} = \Omega_{\text{PBH}} / \Omega_{\text{DM}} \). The PBH density distribution follows that of DM, which is organised into virialised objects (halos) and a diffuse component in the IGM. The number density of PBHs at redshift \( z \) is given by:

\begin{equation}
n_{\text{PBH}}(z) = \frac{\Omega_{\text{DM}} \rho_c f_{\text{PBH}} (1 + z)^3}{M_{\text{PBH}}} = n_{\text{PBH, IGM}}(z) + n_{\text{PBH, halos}}(z),
\end{equation}
where $\rho_c = 3H_0^2/(8\pi G)$ is the critical density of the Universe at \( z=0 \), with \( G = 6.674 \times 10^{-8} \, \text{cm}^3 \, \text{g}^{-1} \, \text{s}^{-2} \) being the gravitational constant. \( M_{\text{PBH}} \) denotes the monochromatic mass of PBHs, while \( n_{\text{PBH, IGM}} \) and \( n_{\text{PBH, halos}} \) represent their number densities in the IGM and within DM halos, respectively.

\begin{itemize}
    \item{PBHs in the IGM:} The number density of PBHs in IGM can be expressed as:

\begin{equation}\label{eq:nPBH_IGM}
n_{\text{PBH}}^{\text{IGM}}(z) = \frac{\Omega_{\text{DM}} \rho_c (1 + z)^3 \left(1 - f_{\text{coll}}(z)\right) f_{\text{PBH}}}{M_{\text{PBH}}},    
\end{equation}

where \( f_{\text{coll}}(z) \) is the fraction of DM that has collapsed into virialised halos by redshift \( z \), calculated using the PS formalism:

\begin{equation}
f_{\text{coll}}(M_{\text{min}}, z) = \text{erfc}\left[\frac{\delta_c(z)}{\sqrt{2\sigma_M^2}}\right],
\end{equation}
with \( \delta_c(z) = 1.686/D(z) \) being the critical density for collapse, \( D(z) \propto (1 + z)^{-1} \) the growth factor, and \( \sigma_M \) the standard deviation of the linearly extrapolated matter power spectrum. \( M_{\text{min}} \) is the minimum halo mass for efficient cooling processes to occur. For \( M_{\text{min}} \), we follow \cite{Liu_2022} and define it as the maximum between (i) the virial mass corresponding to a virial temperature of 100 K, and (ii) the minimum mass required for a halo to host at least one PBH, given its mass and fractional abundance. The 100 K threshold is motivated by the characteristic relative velocity between baryons and DM at \( z \approx 30 \), which corresponds to a baryonic kinetic temperature of approximately 100 K. This velocity suppresses early gas collapse and sets a natural temperature scale for gas capture into halos. Additionally, PBH accretion can heat the surrounding gas above this temperature, reinforcing its use as a conservative lower limit for the onset of significant PBH feedback in halos.

Eq. \ref{eq:nPBH_IGM} indicates that \( n_{\text{PBH}}^{\text{IGM}}(z) \) decreases with decreasing redshift. Besides cosmic expansion, this decrease is attributed to the ongoing structure formation, where more DM collapses into virialised halos, leading to a higher fraction of PBHs being locked into these structures rather than remaining in the IGM.
    \item{PBHs in DM halos:} We assume that the DM distribution within halos follows a Navarro-Frenk-White (NFW) density profile, which is widely used to describe the structure of DM halos. This profile is characterised by a density that decreases with increasing radial distance from the halo centre, described as:

\begin{equation}
\rho_{\mathrm{DM}}(r) = \frac{\rho_c \delta_c}{cx(1 + cx)^2}, 
\end{equation}
where \( x = r / r_{\mathrm{vir}} \) is the radial distance normalised by the virial radius \( r_{\mathrm{vir}} \). The virial radius is expressed as \citep{Barkana_2001}:

\begin{equation}
r_{\mathrm{vir}} = 0.784 \left( \frac{M_{\mathrm{vir}}}{10^8 \, h^{-1} \, ~\mathrm{M_\odot}} \right)^{1/3} 
\left[ \frac{\Omega_m}{\Omega^z_m} \frac{\Delta_c}{18\pi^2} \right]^{-1/3} 
\left( \frac{1 + z}{10} \right)^{-1} h^{-1} \, \mathrm{kpc},
\end{equation}
where the overdensity relative to \(\rho_c\) at the collapse redshift is given by:
\begin{equation}
\Delta_c = 18\pi^2 + 82d - 39d^2,
\end{equation}
with $d = \Omega^z_m - 1$ and $\Omega^z_m = \Omega_m (1 + z)^3/(\Omega_m (1 + z)^3 + \Omega_\Lambda)$.

The parameter \(\Delta_c\) is related to \(\delta_c\) through:

\begin{equation}\label{eq:delta_c}
\delta_c = \frac{\Delta_c}{3} \frac{c^3}{\ln(1 + c) - c / (1 + c)} = \frac{\Delta_c}{3} \frac{c^3}{F(c)}, \tag{8}
\end{equation}
where \( c \) is the concentration parameter, which encapsulates the relationship between the halo's characteristic density and its virial mass. This parameter reflects how tightly matter is bound within the halo, varying with both the virial mass \( M_{\text{vir}} \) and the redshift. It is expressed as:

\begin{equation}
\log c = 1.071 - 0.098 \left[\log \left(\frac{M_{\text{vir}}}{\Msun}\right) - 12\right].
\end{equation}

This relation is based on the N-body simulations by \cite{Maccio_2007}, which are calibrated for halos at \( z = 0 \). Following \cite{Ziparo_2022}, we implement the redshift evolution of the concentration parameter as \( c \propto (1+z)^{-1} \), as suggested by previous works \citep{Barkana_2002, Duffy_2008, Ricotti_2009}.

Within the virial radius of the halo, we can determine the number of PBHs enclosed within any given radial distance $r$, assuming their distribution follows that of the DM. This is expressed as:

\begin{equation}
N_{\text{PBH}}(r) = f_{\text{PBH}} \frac{4\pi}{M_{\text{PBH}}} \int_0^r \rho_{\text{DM}}(r') r'^2 dr',
\end{equation}

This approach allows us to analyse how PBHs are distributed within a DM halo and how their distribution might influence structure formation and radiation emission.
\end{itemize}

\begin{figure}
    \centering
    \includegraphics[width=\linewidth]{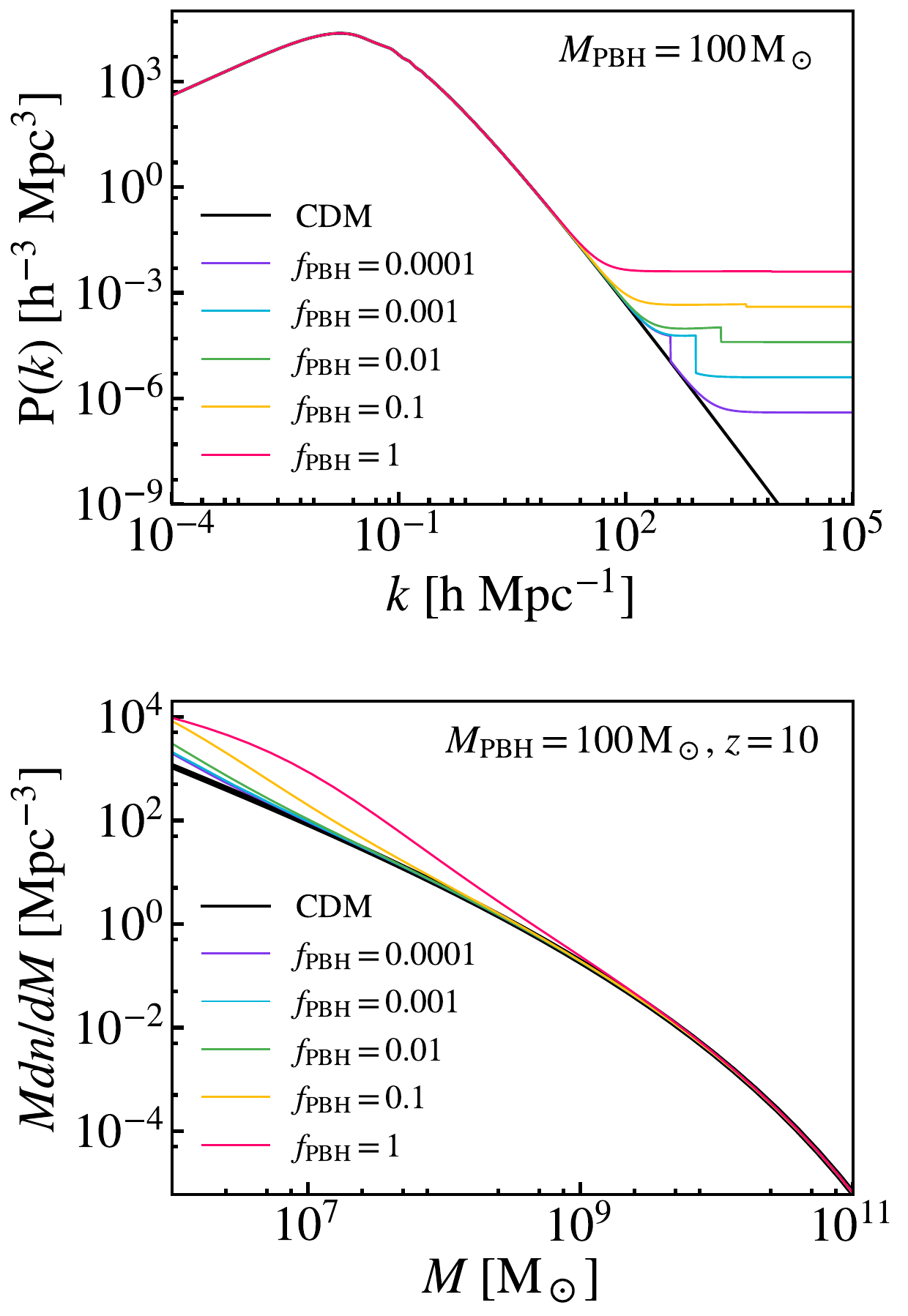}
    \caption{
    Impact of PBHs on the power spectrum and HMF (assuming \(M_{\mathrm{PBH}} = 100 \, \rm{M}_\odot\)). Top panel: The total power spectrum \(P_{\mathrm{tot}}(k)\) as a function of wavenumber \(k\) for various fractions of DM in PBHs (\(f_{\mathrm{PBH}} = 0.0001\), \(0.001\), \(0.01\), \(0.1\), and \(1\)), compared to the standard \(\Lambda\)CDM case (\(f_{\mathrm{PBH}} = 0\)). Bottom panel: The HMF at \(z = 10\), showing the number density of halos as a function of halo mass \(M\), also assuming \(M_{\mathrm{PBH}} = 100 \, \rm{M}_\odot\).
    }
    \label{fig:PS_HMF}
\end{figure}

\subsubsection{Halo mass function in structure formation with PBHs}\label{subsubsec:HMF}

To model the Halo Mass Function (HMF) in cosmologies that include PBHs, we employ the formalism developed by \citet{Zhang_2024a} (see also \citealt{Padilla_2021}), which extends the standard PS approach \citep{Press_1974} to incorporate the effects of PBHs on the power spectrum of density perturbations. This approach modifies the linear power spectrum to include adiabatic perturbations, isocurvature contributions induced by PBHs, and a correlation term that captures the mode mixing between these components.

The total power spectrum, extrapolated to \(z = 0\), is expressed as:
\begin{equation}
P_{\mathrm{tot}}(k) = P_{\mathrm{ad}}(k) + P_{\mathrm{iso}}(k) + P_{\mathrm{corr}}(k),
\end{equation}
where \(P_{\mathrm{ad}}(k)\) denotes the standard adiabatic power spectrum in \(\Lambda\)CDM cosmology, computed using the Planck 2018 cosmological parameters \citep{Planck_2020}. The term \(P_{\mathrm{iso}}(k)\) represents the contribution from isocurvature perturbations arising due to the discreteness of PBHs. Finally, the correlation term \(P_{\mathrm{corr}}(k)\) captures the interaction between adiabatic and isocurvature modes, which is significant at intermediate scales where the influence of PBHs becomes dominant.

The isocurvature contribution \(P_{\mathrm{iso}}(k)\) is given by:

\begin{equation}
P_{\mathrm{iso}}(k) = \frac{[f_{\mathrm{PBH}} D_0]^2}{\bar{n}_{\mathrm{PBH}}}.
\end{equation}
Here, \(\bar{n}_{\mathrm{PBH}}\) represents the average comoving number density of PBHs, while \(D_0 = D_{\mathrm{PBH}}(a = 1)\) denotes the growth factor of PBH-induced perturbations at \(z = 0\). This growth factor evolves as a function of the scale factor \(a\) and can be well approximated using the analytical formalism presented in \cite{Inman_2019}:
\begin{equation}
\begin{aligned}
D_{\text{PBH}}(a) &\approx \left( 1 + \frac{3\gamma a}{2a_- - a_{\text{eq}}} \right)^{a_-}, \quad
\gamma = \frac{\Omega_m - \Omega_b}{\Omega_m}, \\
a_- &= \frac{1}{4} \left( \sqrt{1 + 24\gamma} - 1 \right),
\end{aligned}
\end{equation}
where \(a_{\mathrm{eq}} \sim 1/3400\) corresponds to the scale factor at the epoch of matter-radiation equality.

The mode-mixing term \(P_{\mathrm{corr}}(k)\) is expressed as:
\begin{equation}
P_{\mathrm{corr}}(k) = f_{\mathrm{PBH}} D_0^2 \left(\frac{k}{k_{\mathrm{PBH}}}\right)^3 P_{\mathrm{ad}}(k), 
\quad \text{for } k \leq 3k_{\mathrm{PBH}},
\end{equation}
and vanishes for \(k > 3k_{\mathrm{PBH}}\). Here, \(k_{\mathrm{PBH}} = (2\pi^2 \bar{n}_{\mathrm{PBH}})^{1/3}\) is the characteristic scale below which isocurvature effects dominate.

Using this modified power spectrum, the HMF is computed via the PBH-modified PS formalism, incorporating corrections for ellipsoidal collapse as described in \citet{Sheth_1999}. While this formalism provides a practical framework for studying PBH cosmologies, it has well-known limitations. Specifically, it underestimates the abundance of small halos due to the suppression of their formation and growth by nearby PBH-seeded halos \citep{Zhang_2024b}. At the high-mass end, the PS formalism tends to overpredict the abundance of massive halos by up to a factor of 10 for \( f_{\mathrm{PBH}} \gtrsim 0.01 \), as it does not fully account for the nonlinear suppression of halo growth due to PBH-induced structures and their interactions \citep{LiuBromm2023}. Furthermore, this formalism fails to capture the complex nonlinear effects surrounding individual PBHs, as well as the characteristic bimodal distribution of halos, which emerges from the interplay between isocurvature perturbations and PBH-driven structure formation \citep{Zhang_2024b, LiuBromm2023}. As highlighted in \citet{LiuBromm2023} and \citet{Zhang_2024b}, addressing these limitations is crucial for improving our understanding of PBH-induced structure formation. While the PS formalism is strictly valid for modelling the Poisson effect in the linear regime, the seed effect is not explicitly included in the HMFs derived from this formalism. Future work could explore high-resolution simulations and refined analytical frameworks to better capture nonlinear dynamics and PBH-environment interactions, which remain an open challenge in PBH cosmologies.

In the top panel of Fig.~\ref{fig:PS_HMF}, we show the total power spectrum \(P_{\mathrm{tot}}(k)\) for different values of \(f_{\mathrm{PBH}}\), assuming a PBH monochromatic mass of \(M_{\mathrm{PBH}} = 100 \, \rm{M}_\odot\). Compared to higher masses, such as \(M_{\mathrm{PBH}} = 10^6 \rm{M}_\odot\), which are explored in \citet{Zhang_2024b}, the enhancement at small scales (\(k \gtrsim 10^2 \, h \, \mathrm{Mpc}^{-1}\)) due to the isocurvature component is moderate but still present. This enhancement reflects the gravitational influence of PBHs on small-scale density perturbations, with stronger effects for larger values of \(f_{\mathrm{PBH}}\). 

The bottom panel of Fig.~\ref{fig:PS_HMF} presents the resulting HMF at \(z = 10\) for the same set of \(f_{\mathrm{PBH}}\) and assuming \(M_{\mathrm{PBH}} = 100 \, \rm{M}_\odot\). For this PBH mass, the differences relative to the standard \(\Lambda\)CDM case are most prominent for low-mass halos (\(M \lesssim 10^9 \, \rm{M}_\odot\)), where an increase in their abundance is observed with larger \(f_{\mathrm{PBH}}\), driven by the Poisson effect. In contrast, for halos with \(M \gtrsim 10^9 \, \rm{M}_\odot\), the HMF for \(M_{\mathrm{PBH}} = 100 \, \rm{M}_\odot\) converges to the \(\Lambda\)CDM prediction, indicating negligible influence from PBHs of this mass range on larger halos.
As discussed in \citet{Zhang_2024b}, the analytical results presented here differ from their simulation findings, which exhibit a significant enhancement in the abundance of massive halos and a suppression of low-mass halos. The latter is driven by nonlinear dynamics, such as the engulfment of smaller halos by PBH-hosting massive halos. The differences arise because the analytical framework used in this study accounts only for linear Poisson-induced modifications to the HMF and does not incorporate nonlinear processes, including the seed effect and the redistribution of matter around PBHs.

In cases of extreme \(f_{\mathrm{PBH}}\) values, such as \(M_{\mathrm{PBH}} = 10 \, \Msun\) and \(33 \, \Msun\) with \(f_{\mathrm{PBH}} = 1\), and \(100 \, \Msun\) with \(f_{\mathrm{PBH}} = 0.1\) and \(f_{\mathrm{PBH}} = 1\), the integration of the modified HMF over all halos resulted in a total DM density in halos (\(\rho_{\mathrm{DM}}^{\mathrm{halos}}
\)) exceeding the total cosmological DM density (\(\rho_{\mathrm{DM}}^{\mathrm{total}}\)). This outcome arises because the HMF used here incorporates modifications due to PBHs, which significantly alter DM distribution compared to the standard \(\Lambda\)CDM HMF \citep{Zhang_2024b}. These modifications account for the enhanced abundance of low-mass halos seeded by PBHs, especially in scenarios where PBHs dominate the DM composition (\(f_{\mathrm{PBH}} \sim 1\)).

To address this inconsistency, a correction factor was applied to scale the HMF, ensuring that the total DM density in halos matches the cosmological value (\(\rho_{\mathrm{DM}}^{\mathrm{halos}} = \rho_{\mathrm{DM}}^{\mathrm{total}}\)). This adjustment effectively assigns all DM to halos, setting the DM density in the IGM to zero (\(\rho_{\mathrm{DM}}^{\mathrm{IGM}} = 0\)) for these specific configurations, i.e., in a Universe where massive PBHs dominate the DM budget, and halos encapsulate the entirety of the available DM.

The correction factors applied to the HMF are close to unity across all configurations and were only necessary for \(f_{\mathrm{PBH}} = 1\) and for \(M_{\mathrm{PBH}} = 100 \, \Msun\), also for \(f_{\mathrm{PBH}} = 0.1\). In most cases, the factors deviate by less than \(10\) per cent, with the most significant deviation observed in the extreme scenario of \(M_{\mathrm{PBH}} = 100 \, \Msun\) and \(f_{\mathrm{PBH}} = 1\), where the factor remains below \(30\) per cent. As discussed in Section~\ref{subsec:model_variations}, we tested the impact of using the unmodified HMF, defined here as the standard \(\Lambda\)CDM HMF without accounting for the effects of PBHs, instead of the modified one. These tests revealed that the derived constraints on \(f_{\mathrm{PBH}}\) remain largely unaffected, as the corrections are applied primarily in extreme configurations with large fractions of DM composed of PBHs. For lower \(f_{\mathrm{PBH}}\) values, where the constraints are typically established, the differences between the modified and unmodified HMFs are negligible.

It is worth noting that the dominant contribution of PBHs to observable cosmic backgrounds arises from accretion within halos, where the densities are sufficient to sustain significant accretion rates. In contrast, the contribution from PBHs in the IGM is negligible. Consequently, \(\rho_{\mathrm{DM, IGM}} = 0\) in these extreme cases does not impact the main results, as the dominant contributions to cosmic backgrounds are governed by the accretion processes occurring within halos.

\subsection{Accretion and emission from PBHs}
\label{subsec:accretion}

We consider the same PBH accretion model as in CV2024, outlined below.

\subsubsection{Bondi-Hoyle-Lyttleton model}

We assume the Bondi-Hoyle-Lyttleton accretion model will characterise the accretion processes of PBHs in both the IGM and within halos. The accretion rate is given by:

\begin{equation}
\dot{M}=4 \pi r_{\rm B}^2 \tilde{v} \rho=\frac{4 \pi G^2 M^2 n \mu m_{\rm p}}{\tilde{v}^3},
\label{eq:mdot}
\end{equation}
where \( r_{\rm B} = G M/\tilde{v}^2 \) is the Bondi radius, \( n \) is the gas number density, \( \mu \) is the mean molecular weight, for which we adopt \(\mu = 1.22\), corresponding to that of a primordial neutral gas. \( m_{\rm p} = 1.673 \times 10^{-24} \, \text{g} \) is the proton mass, and \( \tilde{v} \equiv (v^2 + c_{\rm s}^2)^{1/2} \) represents the characteristic velocity. Here, \( v \) is the relative velocity of the PBH with respect to the gas, and \( c_{\rm s} \) is the sound speed of the gas.

It is worth noting that, following the prescription of \citet{Takhistov_2022}, our formulation does not include a Bondi-Eigenvalue factor \( \lambda \), which is sometimes introduced in the literature as an ad hoc correction to account for uncertainties in the accretion process. In this approach, the accretion and emission regimes are determined directly from the local physical conditions of the PBH and the surrounding medium, without the need for additional free parameters.

\subsubsection{Condition for accretion disc formation}
\label{subsec:accretion_condition}

To determine if an accretion disc forms around a PBH of mass \( M \), we examine whether the outer edge of the PBH accretion disc, \( r_{\rm out} \), lies within the radius of the innermost stable circular orbit (ISCO). Efficient disc formation is unlikely if the outer edge is within the ISCO radius. The radius of the outer disc edge is approximately given by \citet{Agol_2002}:

\begin{equation}
r_{\mathrm{out}} \simeq 5.4 \times 10^{9} r_{\rm s}\left(\frac{M}{100 \, \rm{M}_{\odot}}\right)^{\frac{2}{3}}\left(\frac{\tilde{v}}{10 \, \mathrm{km/s}}\right)^{-\frac{10}{3}},
\label{eq:r_out}
\end{equation}
where \( r_{\rm s} = 2 G M/c^2 \) is the Schwarzschild radius of the BH, and \( c = 2.998 \times 10^{10} \, \text{cm/s} \) is the speed of light.

For a non-rotating, spherically symmetric PBH (Schwarzschild PBH), the ISCO radius is defined as

\begin{equation}
r_{\text {ISCO }}=3 r_{\rm s}.
\end{equation}
This radius represents the minimum distance at which an object can maintain a stable circular orbit around the PBH.

\subsubsection{Accretion regimes}
\label{subsec:accretion_regimes}

Following \cite{Takhistov_2022}, we consider different accretion regimes characterised by the dimensionless accretion rate \( \dot{m} \). This rate is defined as:

\begin{equation}
\dot{m} = \frac{\dot{M}}{\dot{M}_{\text{Edd}}} = 2.64 \times 10^{-7} \left(\frac{M}{\Msun}\right) \left(\frac{n}{1 \, \text{cm}^{-3}}\right) \left(\frac{\mu}{1}\right) \left(\frac{\tilde{v}}{10 \, \text{km/s}}\right)^{-3},
\end{equation}
where \( \dot{M}_{\text{Edd}} \) is the Eddington accretion rate given by:

\begin{equation}
\dot{M}_{\text{Edd}} = \frac{L_{\text{Edd}}}{\epsilon_0 c^2} = 6.7 \times 10^{-16} \left( \frac{M}{\Msun} \right) \Msun / \text{s},
\end{equation}
with \( \epsilon_0 = 0.1 \) being the radiative efficiency characteristic of thin discs. The radiative efficiency \( \epsilon_0 \) can vary from 0.057 for a non-rotating Schwarzschild BH to 0.42 for an extremal Kerr BH (see e.g. \citealt{Kato_2008}). As in \cite{Takhistov_2022}, we assume \( \epsilon_0 = 0.1 \) as a typical value for thin discs.

In this study, we include both thin discs and ADAFs. We consider the formation of a thin disc when \( \dot{m} \geq 0.07\alpha \), where $\alpha$ is the viscosity parameter. For a more robust analysis, we further divide ADAFs into three distinct sub-regimes based on the accretion rate, following \cite{Takhistov_2022}. Specifically, we consider LHAF for accretion rates \( 0.1\alpha^2 < \dot{m} < 0.07\alpha \), standard ADAF in the range \( 10^{-3}\alpha^2 < \dot{m} < 0.1\alpha^2 \), and eADAF for \( \dot{m} < 10^{-3}\alpha^2 \). Below, we provide a description of each regime, utilising the model detailed by \cite{Takhistov_2022}, which has also been employed in CV2024.

The thin disc formed around PBHs is optically thick, allowing it to efficiently radiate blackbody emission \citep{Shakura-Sunyaev_1973}, which facilitates a comprehensive analytical model. The temperature profile of the disc, beyond the inner region, is given by \citep{Pringle_1981}:

\begin{eqnarray}
T(r) = T_{\rm i}\left(\frac{r_{\rm i}}{r}\right)^{3 / 4}\left[1-\left(\frac{r_{\rm i}}{r}\right)^{\frac{1}{2}}\right]^{\frac{1}{4}},
\end{eqnarray}
where \( r_{\rm i} \) represents the inner radius of the disc, assumed to be the ISCO radius, and

\begin{eqnarray}
T_{\rm i} = \left(\frac{3GM\dot{M}}{8\pi r^3_{\rm i}\sigma_{\rm B}}\right) = 53.3\, \rm{eV} \left(\frac{n}{1\, \rm{cm^{-3}}}\right)^{1/4} \left(\frac{\tilde{v}}{10\, \rm{km/s}}\right)^{-3/4},
\label{eq:T}
\end{eqnarray}
where \( \sigma_{\rm B} = 5.67 \times 10^{-5} \, \text{g/s}^3/\text{K}^4 \) is the Stefan-Boltzmann constant.

Beyond the inner edge, the disc reaches a maximum temperature \( T_{\max} = 0.488 T_{\rm i} \) at a radius \( r = 1.36 r_{\rm i} \).

The thin disc spectrum is a combination of blackbody emissions from different radii. Using the scaling relations from \citet{Pringle_1981} and ensuring continuity, the resulting spectrum can be approximated as:

\begin{eqnarray}
\begin{aligned}
& \nu < T_{\rm o}: \quad L_{\nu} = c_{\alpha} \left(\frac{T_{\max }}{T_{\rm o}}\right)^{5/3} \left(\frac{\nu}{T_{\max }}\right)^{2}, \\
& T_{\rm o} < \nu < T_{\max }: \quad L_{\nu} = c_{\alpha} \left(\frac{\nu}{T_{\max }}\right)^{1/3}, \\
& T_{\max } < \nu: \quad L_{\nu} = c_{\alpha} \left(\frac{\nu}{T_{\max }}\right)^{2} e^{1-\nu/T_{\max }},
\end{aligned}
\end{eqnarray}
where \( \nu \) is the photon energy, \( T_{\rm o} = T(r_{\rm out}) \) is the temperature at the outer edge of the disc, and

\begin{eqnarray}
\begin{aligned}
c_{\alpha}= &1.27 \times 10^{29} \mathrm{erg}\; \mathrm{eV}{ }^{-1} \mathrm{~s}^{-1}\left(\frac{M}{\rm{M}_{\odot}}\right)^{2}\\&\left(\frac{n}{1 \mathrm{~cm}^{-3}}\right)^{3 / 4}\left(\frac{\tilde{v}}{10 \mathrm{~km} / \mathrm{s}}\right)^{-9 / 4}.
\end{aligned}
\end{eqnarray}
\( c_{\alpha} \) is normalised such that the emission achieves the maximum possible efficiency for a Schwarzschild BH (\( \int L_{\nu} \, d\nu = 0.057 \dot{M} c^{2} \)).

In contrast to the thin disc, when an ADAF forms, the heat generated by viscosity is not efficiently radiated away, and a significant amount of energy is advected into the BH event horizon along with the gas. This results in a complex emission spectrum with contributions from electron cooling, synchrotron radiation, inverse-Compton (IC) scattering, and bremsstrahlung processes.

To characterise the flow, we use the following parameters: the fraction of viscously dissipated energy that heats electrons directly, \( \delta = 0.3 \); the ratio of gas pressure to total pressure, \( \beta = 10/11 \); the minimum flow radius, equal to the ISCO radius, \( r_{\min} = 3 r_{\rm s} \); and the viscosity parameter, \( \alpha = 0.1 \). These values are consistent with recent simulations and observations \citep{Yuan_2014}, and are also used in the model by \citet{Takhistov_2022}.

The synchrotron emission is self-absorbed and peaks at a photon energy of

\begin{eqnarray}
\begin{aligned}
\nu_p= & 1.83 \times 10^{-2} \mathrm{eV}  \left(\frac{\alpha}{0.1}\right)^{-\frac{1}{2}}\left(\frac{1-\beta}{1 / 11}\right)^{\frac{1}{2}} \\& \theta_{\rm e}^2\left(\frac{r_{\min }}{3 r_{\rm s}}\right)^{-\frac{5}{4}}\left(\frac{M}{\rm{M}_{\odot}}\right)^{-\frac{1}{2}}\left(\frac{\dot{m}}{10^{-8}}\right)^{\frac{3}{4}},
\end{aligned}
\end{eqnarray}
where \( \theta_{\rm{e}} \) represents the temperature in units of the electron mass \( m_{\rm{e}} \),

\begin{eqnarray}
\theta_{\rm e} = \frac{k T_{\rm e}}{m_{\rm e} c^2} = \frac{T_{\rm e}}{5.93 \times 10^9 \, \text{K}}.
\end{eqnarray}
where $k_{\rm{B}}=1.3807\times10^{-16}$ cm$^{2}$ g s$^{-2}$ K$^{-1}$ is the Boltzmann constant and $T_{\rm{e}}$ the electron temperature. The peak luminosity is given by

\begin{eqnarray}
L_{\rm{\nu_p}} = 5.06 \times 10^{38} \, \text{erg/s/eV} \, \alpha^{-1} (1-\beta) \left(\frac{M}{\Msun}\right) \dot{m}^{3/2} \theta_{\rm e}^5 \left(\frac{r_{\text{min}}}{r_{\rm{s}}}\right)^{-1/2}.
\end{eqnarray}

In this simple description, the synchrotron spectrum is assumed to terminate at \( \nu_p \), resulting in \( L_{\nu, \mathrm{syn}} = 0 \) for \( \nu > \nu_p \) \citep{Mahadevan_1997}.

The synchrotron photons undergo IC scattering with the surrounding electron plasma. The resulting IC spectrum is described by

\begin{eqnarray}
L_{\nu, \mathrm{IC}} = L_{\rm \nu_p} \left(\frac{\nu}{\nu_p}\right)^{-\alpha_c},
\end{eqnarray}
where

\begin{eqnarray}
\alpha_{\rm c} = -\frac{\ln \tau_{\rm e s}}{\ln A},
\end{eqnarray}
with \( \tau_{\rm e s} = 12.4 \dot{m} \alpha^{-1} \left(\frac{r_{\text{min}}}{r_{\rm{s}}}\right)^{-1/2} \) as the electron scattering optical depth, and \( A = 1 + 4 \theta_{\rm e} + 16 \theta_{\rm e}^2 \). This expression for $L_{\nu, \mathrm{IC}}$ is valid only in the frequency range $\nu_p \leq \nu \lesssim 3 k_B T_e $, which corresponds to the energy range where thermal IC scattering significantly modifies the photon spectrum. Beyond this range, photons either lack sufficient energy for further upscattering or are dominated by other emission processes.

Finally, the bremsstrahlung emission from the thermal spectrum is given by

\begin{equation}
\begin{aligned}
L_{\nu, \text { brems }}= & 1.83 \times 10^{17}\left(\frac{\alpha}{0.1}\right)^{-2}\left(\frac{c_1}{0.5}\right)^{-2} \ln \left(\frac{r_{\max }}{r_{\min }}\right) F\left(\theta_{\rm e}\right) \\
& \left(\frac{T_{\rm e}}{5 \times 10^9 \mathrm{~K}}\right)^{-1} e^{-\left(h \nu / k T_{\rm e}\right)} \left(\frac{M}{\rm{M}_{\odot}}\right) \dot{m}^2\;\mathrm{ergs}\;\mathrm{s}^{-1} \mathrm{~Hz}^{-1},
\end{aligned}
\end{equation}
where \( c_1 = 0.5 \) and \( F(\theta_{e}) \) is given by \citet{Mahadevan_1997}:

\begin{equation}
F(\theta_{e}) =
\begin{cases}
\begin{aligned}
& 4\left(\frac{2 \theta_{e}}{\pi^{3}}\right)^{1/2}\left(1+1.78 \theta_{e}^{1.34}\right) \\
& + 1.73 \theta_{e}^{3/2}\left(1+1.1 \theta_{e}+\theta_{e}^{2}-1.25 \theta_{e}^{5/2}\right),
\end{aligned}
& \theta_{e} \leq 1, \\
\begin{aligned}
& \left(\frac{9 \theta_{e}}{2 \pi}\right)\left(\ln \left[1.12 \theta_{e}+0.48\right]+1.5\right) \\
& + 2.30 \theta_{e}\left(\ln \left[1.12 \theta_{e}\right]+1.28\right),
\end{aligned}
& \theta_{e} \geq 1.
\end{cases}
\end{equation}

Regarding frequency ranges, synchrotron emission dominates the radio to submillimeter spectrum due to its origin in the thermal electron population of the ADAF. The IC scattering process primarily contributes to the X-ray region, with photon energies typically extending up to \( \sim 3 k_B T_e \). This upper limit corresponds to the maximum energy gain for photons undergoing thermal IC scattering before other emission mechanisms become dominant. Finally, bremsstrahlung emission is significant in the hard X-ray region, where it provides a relatively constant contribution across a wide frequency range before an exponential cutoff dictated by the high-energy tail of the electron distribution.

Each ADAF subregime has different temperature dependences because \( \theta_{\rm{e}} \) is determined by balancing heating and radiative processes \citep{Mahadevan_1997}. Direct viscous electron heating dominates at low accretion rates, while ion-electron collisional heating becomes significant at higher rates. The methodology for calculating \( \theta_{\rm{e}} \) in each subregime can be found in Section 3.1.3 of \cite{Takhistov_2022}.

We acknowledge that the emission model adopted in this work, following \citet{Takhistov_2022}, involves several simplifying assumptions that may limit its physical completeness. These include the use of fixed microphysical parameters—such as the viscosity parameter \( \alpha \), the plasma \( \beta \) (which reflects the relative importance of gas pressure over magnetic pressure), and the electron heating fraction \( \delta \)—which are kept constant across all accretion regimes where they apply. The model also considers only thermal electron populations and does not include non-thermal particles, which could affect the emitted spectrum in some ADAF scenarios. Furthermore, the emission is computed assuming spherical symmetry in the geometry of the accretion flow, neglecting potential effects of angular momentum and anisotropies in the gas surrounding PBHs. While the slim disc regime, relevant at high Eddington ratios, is not included in this model, we explicitly verified that its omission does not affect our results: the super-Eddington regime is only marginally reached in the densest central regions of the most massive halos, which contribute negligibly to the total background. These assumptions are standard in the ADAF literature and enable a tractable and physically motivated framework, but they may introduce systematic uncertainties that cannot be robustly quantified within the scope of this work. We highlight these caveats as limitations of the current implementation and note that they warrant further investigation.

\begin{figure*}
    \centering
    \includegraphics[width=\textwidth]{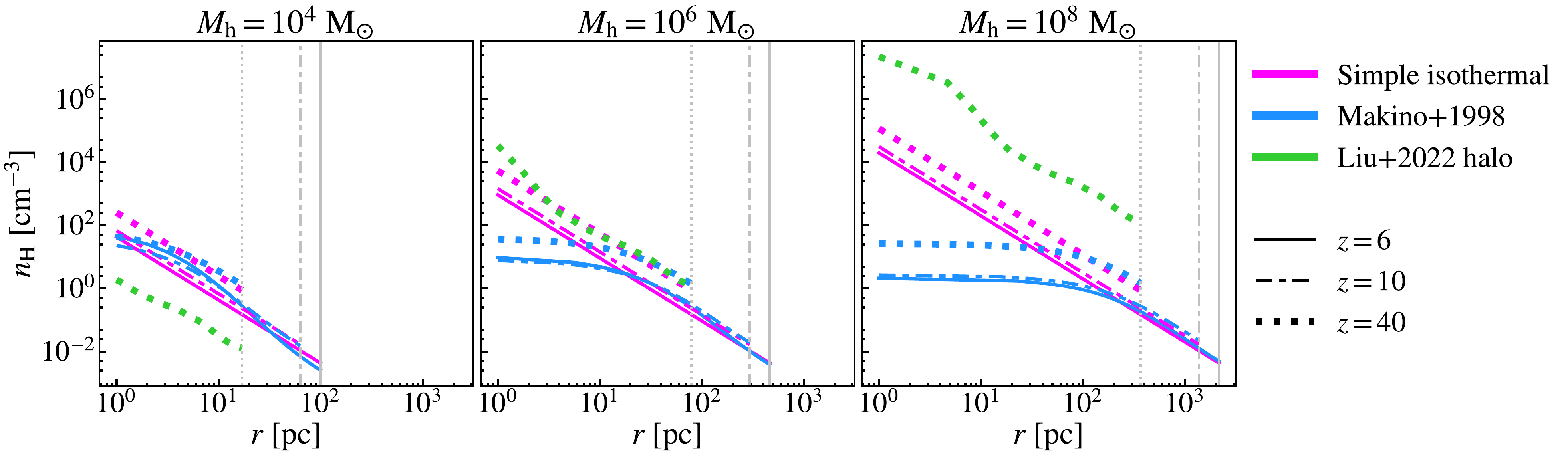}
    \caption{Comparison of hydrogen number density profiles $n_{\text{H}}(r)$ within DM halos of different masses ($M_{\text{h}} = 10^4\;\Msun$, $10^6\;\Msun$, and $10^8\;\Msun$) at redshifts $z = 6$ (solid lines), $z = 10$ (dashed-dotted lines), and $z = 40$ (dotted lines). The three density profiles examined include a simple isothermal profile (magenta), the profile derived by \cite{Makino_1998} (blue), and the rescaled profile from \cite{Liu_2022} (green). The vertical grey lines denote the virial radii of the corresponding halos at each redshift.}
    \label{fig:density_profiles}
\end{figure*}

\subsection{PBH accretion in halos}\label{sec:halo_accretion}

As discussed earlier, the accretion rate of PBHs depends not only on the PBH mass but also on the characteristic velocity of the surrounding medium and the gas density. Within halos, we assume that the virial velocity of the halos determines the characteristic velocity:

\begin{equation}
\tilde{v} \approx \sqrt{\frac{G M_{\rm{h}}}{r_{\text{vir}}}} \sim 5.4 \; \text{km}\; \text{s}^{-1} \left( \frac{M_{\rm{h}}}{10^6\; \Msun} \right)^{-1/2} \left( \frac{1 + z}{21} \right)^{-1/2}.
\end{equation}

Regarding the gas density, we consider three distinct halo density profiles to assess their impact on the calculated radiation: 

\begin{itemize}
    \item Simple isothermal density profile. The first density profile considered is a simple isothermal profile, where the density distribution follows an inverse-square law given by:

\begin{equation}
n(r) = \frac{n_0}{3} \left(\frac{r_{\text{vir}}}{r}\right)^2,
\end{equation}
where \( n_0 \) is the average gas density, defined as

\begin{equation}
n_0 = 200 \rho_{\rm{m}}(a) \frac{\Omega_{\rm{b}}}{\Omega_{\rm{m}}} \frac{1}{\mu m_{\rm{p}}},
\end{equation}
with \(\rho_{\rm{m}}(a)\) the matter density of the Universe as a function of the scale factor \( a \), given by

\begin{equation}
\rho_{\rm{m}}(z) = \frac{3\Omega_{\rm{m}} H_0^2}{8 \pi G} (1+z)^3,
\end{equation}

This profile assumes a spherically symmetric distribution where the density falls off as \( r^{-2} \), providing a simple and widely used model in astrophysical studies. Simulations indicate that the gas distribution in high-\(z\) atomic-cooling halos approximately follows \(\rho \propto r^{-2}\) at \(r \gtrsim 0.003 \, \text{pc} \) \citep{Safarzadeh_Haiman2020}. Similar density profiles are observed in the simulations by \cite{Liu_2022} for molecular-cooling minihalos. Consequently, \cite{Liu_2022} employs this profile for their calculation of the radiation background produced by PBHs.
    \newline
    \item Makino+1998 density profile. The second density profile is the one derived in \cite{Makino_1998}. The gas density profile in hydrostatic equilibrium within the potential of a DM halo, assuming an isothermal distribution, can be approximated by the $\beta$-model:

\begin{equation}
\rho_{\text{gas}} = \rho_0 \exp \left\{-\frac{\mu m_{\rm{p}}}{2 k_{\rm{B}} T_{\text{vir}}} \left[v_e^2(0) - v_e^2(r)\right]\right\},
\end{equation}
where \( \rho_0 \) is the central gas density, 
\begin{equation}
\rho_{0}(z) = \frac{(\Delta_c / 3) c^3 e^A}{\int_{0}^{c} (1 + t)^{A / t} t^2 dt} \left( \frac{\Omega_{\rm{b}}}{\Omega_{\rm{m}}} \right) \Omega_{\rm{m}}^z \rho_{\text{c}}(z),
\end{equation}
where \( A = 2c/F(c) \), with \( F(c) \) defined in Eq.~\ref{eq:delta_c}, and \(\rho_{\text{c}}(z) = 3H(z)^2/(8\pi G)\) represents the critical density of the Universe at a redshift \(z\).

The viral temperature of the halo, \( T_{\text{vir}} \), is given by:

\begin{equation}
T_{\text{vir}} = 1.98 \times 10^4 \left(\frac{\mu}{0.6}\right) \left(\frac{M_{\text{vir}}}{10^8 h^{-1} \Msun}\right)^{2/3} \left[\frac{\Omega_{\rm{m}} \Delta_c}{\Omega_{\rm{m}}^z 18 \pi^2}\right]^{1/3} \left(1 + z \right) \, \text{K},
\end{equation}
and \( v_e \) is the escape velocity

\begin{equation}
v_e^2(r) = 2 \int_r^{\infty} \frac{GM(r')}{r'^2} dr' = 2 v_{c,\text{vir}}^2 \frac{F(cx) + \frac{cx}{1+cx}}{xF(c)},
\end{equation}
where \( v_{c,\text{vir}} = \sqrt{GM_{\text{vir}}/r_{\text{vir}}} \) is the virial velocity.

These equations collectively describe how the gas density profile is determined in the context of a universal DM halo, accounting for the effects of hydrostatic equilibrium and the influence of the DM distribution. \cite{Ziparo_2022} use this profile to calculate the radiation backgrounds produced by PBHs.
\newline
\item Rescaled halo from Liu+2022. The third density profile is derived by rescaling a halo from the study by \cite{Liu_2022}. In this model, the density profile at the onset of star formation at \( z \approx 30 \) within a halo of \( 2 \times 10^5 \, \rm{M}_{\odot} \) is utilised. The profile is rescaled to match the virial mass of the halos in our study by multiplying the reference profile by the ratio of the enclosed mass within our halos at their respective virial radii to the enclosed mass of the reference halo at its virial radius. This method was also employed in CV2024 to address resolution issues and to better account for the density distribution within halos at $z\approx30$.

This profile is only employed up to \(z = 20\), beyond which its applicability diminishes. At lower redshifts, hierarchical mergers, baryonic feedback processes (e.g., supernova-driven outflows and radiation pressure), and longer dynamical timescales significantly alter halo density structures. These effects deviate from the assumptions underlying the rescaled profile derived from a halo at \(z \sim 30\). Limiting its use to \(z = 20\) ensures the validity of the model within the appropriate dynamical and physical regimes.

\end{itemize}

As illustrated in Fig. \ref{fig:density_profiles}, the hydrogen number density profiles exhibit distinct behaviours depending on the chosen model and halo mass, with significant variations observed across different redshifts. These differences increase with halo mass as the density profiles deviate more substantially from one another at higher masses. However, we calculated that the contribution of halos with \(M_{\text{h}} > 10^6 \, M_\odot\) to the total emission is negligible, as halos with \(M_{\text{h}} < 10^6 \, M_\odot\) dominate the total emission, contributing the overwhelming majority of the radiation.

\subsection{PBH accretion in the IGM}\label{sec:IGM_accretion}

Following \cite{Ricotti_2008b}, we assume that PBHs in the IGM are surrounded by a uniformly distributed gas with a density given by:

\begin{equation}
\rho_{\text{IGM}} = 250 \mu m_{\rm{p}} \left( \frac{1 + z}{1000} \right)^3 \, \text{g} \, \text{cm}^{-3}.
\end{equation}
This relation provides the mean gas density in the IGM under the assumption of a homogeneous and isotropic Universe. As discussed by \cite{Ricotti_2008b}, the scaling $\rho_{\text{IGM}} \propto (1+z)^3$ reflects the cosmic expansion, where the gas density evolves inversely with the comoving volume. The prefactor $250 \mu$ corresponds to the baryon density at $z = 1000$, with $\mu$ accounting for the primordial composition of the gas, dominated by hydrogen and helium. This density serves as a crucial input for modelling gas dynamics and accretion processes around PBHs. 

While we follow the assumption of a uniform gas distribution in the IGM, as adopted in \cite{Ziparo_2022}, it is important to acknowledge that the presence of DM halos seeded by PBHs could concentrate the surrounding gas. This effect may enhance accretion rates and luminosity, potentially influencing feedback processes and the thermal evolution of the IGM. However, our constraints indicate that the fraction of DM composed of PBHs is quite small in order to remain consistent with the observational and theoretical limits explored in this work, suggesting that this effect is probably not important.

The sound speed of the gas, $c_s$, is determined using the parametric relation provided by \cite{DeLuca_2020}, which incorporates the evolution of the CMB temperature with redshift. This expression is given by:

\begin{equation}
c_s \simeq 5.70 \left( \frac{1 + z}{1000} \right)^{1/2} 
\left[ \left( \frac{1 + z_{\text{dec}}}{1 + z} \right)^{\beta} + 1 \right]^{-1/(2\beta)} \, \text{km s}^{-1},
\end{equation}
where $\beta = 1.72$ is a fitting parameter, and $z_{\text{dec}} \simeq 130$ corresponds to the redshift at which baryons decouple from the radiation field. This formulation captures the transition from tightly coupled baryons and radiation at early times to a thermally evolving IGM. The inclusion of this relation ensures an accurate representation of the baryonic sound speed across cosmic epochs, which is critical for assessing gas dynamics and the efficiency of accretion onto PBHs.

To model the relative velocity between DM and baryons, we adopt the relation proposed by \cite{Ali-Haimoud_Kamionkowski_2017}:

\begin{equation}
v_{\text{rel}}(z) \simeq 30 \min \left[ 1, \left( \frac{1 + z}{1000} \right) \right] \, \text{km} \, \text{s}^{-1}.
\end{equation}

This residual streaming velocity originates after the recombination epoch, when baryons kinetically decouple from photons at $z \sim 1000$. The root-mean-square (RMS) value of this velocity is approximately 30 km/s at recombination, as calculated by \cite{Ali-Haimoud_Kamionkowski_2017}. The velocity decreases as $(1+z)$ at earlier times, reflecting the expansion of the Universe, and remains constant for $z \lesssim 1000$. This streaming motion is crucial for understanding the thermal and dynamical evolution of the IGM, as it affects small-scale structure formation and the interactions between baryons and DM \citetext{see also \citealt{Kashlinsky_2021,Atrio-Barandela_2022}}.

\subsection{Radiation backgrounds}\label{subsubsec:rad_backgrounds}

The radiation background produced by PBHs accreting within halos and the IGM is analysed across three frequency regimes: X-ray, LW, and radio. This section outlines the equations used to characterise the background intensity in each regime, while a detailed derivation is provided in Appendix~\ref{appendix:derivation}. 

We acknowledge that the Cosmic Infrared Background (CIB) has also been studied as a potential probe of PBHs. Some works have examined whether accreting PBHs could account for the observed CIB–CXB cross-correlation \citep[e.g.,][]{Kashlinsky_2016, Cappelluti_2017,Cappelluti_2022}. More recent analyses incorporating detailed accretion physics and thermal feedback suggest that PBHs are subdominant contributors. For example, \citet{Hasinger_2020}, assuming a broad PBH mass function extending from $10^{-8}$ to $10^{10}\,M_\odot$, Bondi-like accretion from homogeneous background gas, and a PBH abundance equal to the total DM, estimated a contribution of only $\sim$0.5\% to the CIB fluctuations. \citet{Manzoni_2024}, considering both halo and background-gas accretion with a self-consistent treatment of thermal evolution, evaluated PBH contributions to the CIB for both monochromatic and lognormal mass functions in the range $1$–$10^3\,\rm{M}_\odot$. In the limiting case where PBHs constitute all of the DM, they found a contribution below 1\% to the observed NIRB fluctuations. This value drops to below 0.1\% when current observational constraints on PBH abundance are taken into account, with a maximal contribution at $M_{\mathrm{PBH}} \sim 50\,\rm{M}_\odot$ within the allowed parameter space. Given this, and our focus on radiation backgrounds that place more direct constraints on the PBH contribution to DM, we do not consider the CIB further in this work.

The comoving emissivity is calculated differently for halos and the IGM. In the case of halos, integration over the HMF is required, whereas for the IGM, the emissivity is determined directly from the PBH number density.

For halos, the comoving emissivity within a specific frequency band $\nu\in[\nu_1, \nu_2]$ is given by:

\begin{equation}
    \epsilon_{\rm cm,[\nu_1, \nu_2]}(z) = \int L_{[\nu_1, \nu_2]}(M_{\rm h},z) \frac{dn_{\rm h}}{dM_{\rm h}} dM_{\rm h},
\end{equation}
where $ L_{[\nu_1, \nu_2]}(M_{\rm h},z) $ represents the total luminosity of a halo with mass $ M_{\rm h} $ at redshift $ z $. This luminosity is computed using the accretion model described in Section~\ref{subsec:accretion_regimes}, incorporating the gas density profiles and characteristic velocities discussed in Section~\ref{sec:halo_accretion}. The term $ dn_{\rm h}/dM_{\rm h} $ corresponds to the HMF, which provides the comoving number density of halos per unit halo mass, further detailed in Section~\ref{subsubsec:HMF}.

For PBHs in the IGM, the comoving emissivity is expressed as:

\begin{equation}
   \epsilon_{\rm cm,[\nu_1, \nu_2]}(z) = L_{[\nu_1, \nu_2]}(z) n_{\text{IGM}}(z),
\end{equation}
where $ L_{[\nu_1, \nu_2]}(z) $ denotes the total luminosity per PBH within the band $[\nu_1, \nu_2]$ in the IGM, and $ n_{\text{IGM}}(z) $ represents the comoving PBH number density, as defined in Section~\ref{subsubsec:distribution}. As with halos, the luminosity is computed using the formalism in Section~\ref{subsec:accretion_regimes}, with gas density and velocity profiles specific to IGM conditions, discussed in Section~\ref{sec:IGM_accretion}.

The background intensity in the X-ray band, observed at $ z=0 $, is obtained by discretising the redshift integration to account for the contributions $\Delta J_{\mathrm{X},i}$ of individual redshift bins:

\begin{equation}
    J_{\rm X} = \sum_{i}\Delta J_{\mathrm{X},i} = \sum_{i} \frac{1}{4\pi} \frac{\epsilon_{\rm cm,X}(z_i)}{(1+z_i)^2} \Delta r_i,
    \label{eq:JX}
\end{equation}
where $ \epsilon_{\rm cm,X}(z_i) $ is the comoving X-ray emissivity at redshift $ z_i $, and $ \Delta r_i $ represents the comoving distance step.

For the LW band, the integrated physical intensity observed at $ z $ is first calculated as:

\begin{equation}
    J_{\rm LW,int}(z) = \frac{(1+z)^4}{4\pi} \int_{z}^{z_{\max}(z)} \frac{\epsilon_{\rm cm,LW}(z,z')}{(1+z')^3} c \left. \frac{dt_{\rm U}}{da'} \right|_{z'} dz',
    \label{eq:JLW}
\end{equation}
where $ t_{\rm U} $ is the cosmic age and $ z_{\max}(z) = (13.6/11.2)(1 + z) - 1 $, which corresponds to the maximum redshift from which photons originally emitted at 13.6 eV are redshifted to 11.2 eV at $ z $, marking the relevant range for the LW band. The corresponding dimensionless specific intensity is given by:

\begin{equation}
    J_{\rm LW}(z) = \frac{10^{21} J_{\rm LW,int}(z) / \Delta \nu_{\rm LW}}{\text{erg} \ \text{s}^{-1} \ \text{cm}^{-2} \ \text{sr}^{-1} \ \text{Hz}^{-1}},
\end{equation}
where $ \Delta \nu_{\rm LW} = (13.6 - 11.2) \ \text{eV} / h_{\rm P}$, with $h_{\rm P} = 4.1357 \times 10^{-15} \ \text{eV s}$ being the Planck constant.

For the radio band, the physical background intensity at an observed frequency $\nu$ is computed as:

\begin{equation}
    J_{\nu} = \frac{1}{4\pi} \int_{z}^{\infty} \frac{(1+z)^3}{(1+z')^2} 
    \dot{\rho}_{\nu'}(z') c\left. \frac{dt_{\rm U}}{da'} \right|_{z'}
dz'.
\label{eq:Jnu}
\end{equation}
where $ \dot{\rho}_{\nu'}(z') \equiv \epsilon_{\nu'}(z')/(1+z')^3 $ represents the comoving emissivity at source-frame frequency $\nu'=\nu[(1+z')/(1+z)]$. The corresponding brightness temperature, which is a crucial observable for 21 cm cosmology, is calculated as:

\begin{equation}
    T_b(\nu, z) = \frac{J_{\nu}(z) c^2}{2 k_B \nu^2}.
\end{equation}

\section{Results and discussion}\label{sec:results}

In this section, we analyse the contribution of PBHs to the CXB, LWB, and CRB, exploring their potential role as significant sources of these cosmic backgrounds and as candidates for DM. Using theoretical predictions for different PBH masses and fractions, we compare their impact with current observational constraints. The analysis considers contributions from both halos and the IGM, highlighting the interplay between these components in shaping the radiation backgrounds. Finally, we discuss the implications of these results for the thermal and ionisation history of the Universe, the suppression of star formation, and the constraints they impose on the PBH parameter space.

\subsection{Background intensity}
\subsubsection{X-ray background}\label{subsubsec:xray}

Approximately 70 per cent of the CXB is attributed to resolved sources, predominantly AGN, across both soft and hard X-ray bands \citep{Cappelluti_2017, Cappelluti_2022}. The unresolved fraction of the CXB is likely due to a yet undetected population of BHs, such as those accreting in heavily obscured environments \citep{Gilli_2007, Treister_2009}, as well as extended X-ray emissions from galaxy clusters \citep{Gilli_1999}. In addition, there could be a contribution from high-$z$ AGN, linked to the recent JWST discovery of such abundant sources, but with significant uncertainties linked to the physics of their X-ray emission \citetext{e.g., \citealt{Jeon_2022,Maiolino_2024}}.

To investigate the potential contribution of PBHs to this unresolved background, we quantify their impact on the X-ray background in the soft (0.5 -- 2 keV) and hard (2 -- 10 keV) X-ray bands. Our analysis specifically aims to determine whether PBHs can account for a significant portion of the unresolved CXB, thereby offering insights into the nature of DM.

\begin{figure*}
    \centering
    \includegraphics[width=\textwidth]{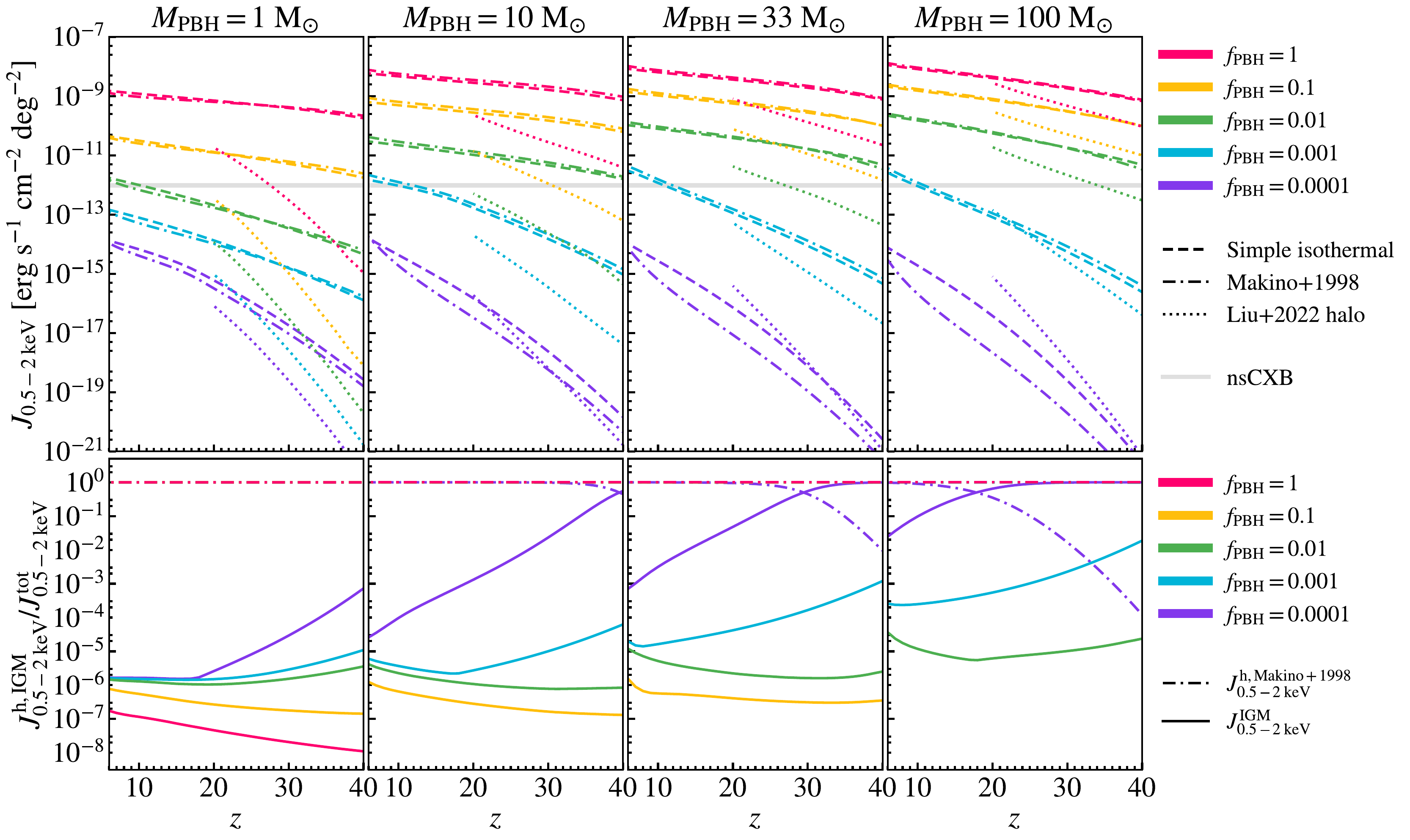}
    \caption{ 
        Contribution of accreting high-$z$ BH sources to the present-day ($z=0$) soft X-ray background (0.5–2 keV). The top row shows the cumulative evolution of the integrated background intensity, $J_\mathrm{X}$, as a function of redshift $z$, for PBH masses of $M_{\text{PBH}} = 1 \, \Msun$, $10 \, \Msun$, $33 \, \Msun$, and $100 \, \Msun$ in each column. Different values of $f_{\text{PBH}}$ are represented by distinct colours, as indicated in the legend, ranging from $10^{-4}$ to 1. The grey shaded region corresponds to the observed excess X-ray background reported in \cite{Cappelluti_2017}. Three halo density profiles are considered: simple isothermal (dashed lines), the profile from \cite{Makino_1998} (dashed-dotted lines), and the profile from \cite{Liu_2022} (dotted lines). The bottom row illustrates the relative contributions of halos ($J^{\text{h}}_{\text{X}}$) and the IGM ($J^{\text{IGM}}_{\text{X}}$) to the total X-ray background intensity ($J^{\text{tot}}_{\text{X}}$). These are represented as $J^{\text{h}}_{\text{X}}/J^{\text{tot}}_{\text{X}}$ (dashed-dotted lines) and $J^{\text{IGM}}_{\text{X}}/J^{\text{tot}}_{\text{X}}$ (solid lines), respectively. The relative contributions are computed using the same PBH masses and density profile configurations as in the top row.}
    \label{fig:soft_Xray}
\end{figure*}

The results in Fig. \ref{fig:soft_Xray} illustrate the cumulative contribution of accreting PBHs to the present-day soft X-ray background intensity, \( J_{0.5-2\rm{keV}} \), for various PBH masses and fractions \( f_{\text{PBH}} \). The shaded grey region indicates the observed excess X-ray background reported by \citet{Cappelluti_2017}, which serves as a constraint on unresolved contributions to the CXB.

The analysis shows that the PBH contribution depends significantly on both the mass and the fraction \( f_{\text{PBH}} \). For the simple isothermal and Makino+1998 density profiles, PBHs with \( M_{\text{PBH}} = 1 \, \Msun \) are ruled out for \( f_{\text{PBH}} \geq 0.01 \), while for \( M_{\text{PBH}} = 10 \, \Msun \), \( 33 \, \Msun \), and \( 100 \, \Msun \), fractions \( f_{\text{PBH}} \geq 0.001 \) are excluded due to their overproduction of soft X-ray background intensity. The results for the simple isothermal and Makino+1998 profiles are fully consistent across all cases.

The results based on the rescaled halo profile from \citet{Liu_2022} are shown only up to \(z = 20\), due to the redshift limitations of the model, as discussed in Section~\ref{sec:halo_accretion}. Within this range, the constraints are less restrictive compared to the simple isothermal and Makino+1998 profiles. For \( M_{\text{PBH}} = 1 \, \Msun \), fractions \( f_{\text{PBH}} = 1 \) are excluded, while for \( M_{\text{PBH}} = 10 \, \Msun \), \( 33 \, \Msun \), and \( 100 \, \Msun \), fractions \( f_{\text{PBH}} \geq 0.1 \), \( f_{\text{PBH}} \geq 0.01 \), and \( f_{\text{PBH}} \geq 0.01 \), respectively, are ruled out.

The differences in constraints between the Liu+2022 halo profile and the other two profiles stem from variations in the predicted gas densities across different halo masses. For halos with \(M_{\text{h}} < 5 \times 10^5 \, \Msun\), the Liu+2022 profile predicts lower gas densities compared to the simple isothermal and Makino+1998 profiles, leading to reduced X-ray emissions from these low-mass halos. Since halos in this mass range dominate the total X-ray luminosity due to their abundance, as discussed in Section~\ref{sec:halo_accretion}, this generally results in lower overall X-ray output for the Liu+2022 profile.

However, the Liu+2022 profile predicts higher gas densities for halos with \(M_{\text{h}} \geq 5 \times 10^5 \, \Msun\), and this difference becomes relevant for certain configurations. Specifically, for higher PBH masses (\(M_{\text{PBH}} = 33 \, \Msun\) and \(M_{\text{PBH}} = 100 \, \Msun\)) and very low fractions (\(f_{\text{PBH}} = 10^{-4}\)), the increased contribution from halos in the intermediate mass range (\(5 \times 10^5 \, \Msun \leq M_{\text{h}} \leq 1 \times 10^6 \, \Msun\)) leads to slightly higher X-ray emissions for the Liu+2022 profile compared to the other two profiles.

This behaviour reflects a balance between the higher gas densities of intermediate-mass halos and the greater abundance of low-mass halos. For small fractions of PBHs (\(f_{\text{PBH}} = 10^{-4}\)), the number of PBHs increases significantly, amplifying the contribution from the intermediate-mass halos where the Liu+2022 profile predicts higher densities. This interplay results in slightly higher emissions for the Liu+2022 profile in these specific cases, despite its generally lower predictions for the total X-ray output. This distinction highlights the sensitivity of PBH constraints to the choice of halo density profile and the complex role of halo mass distribution in shaping X-ray emissions.

The bottom panels of Fig.~\ref{fig:soft_Xray} illustrate the relative contributions of PBHs in halos and the IGM to the total soft X-ray intensity. In general, the contribution from the IGM decreases with decreasing redshift, while the contribution from halos becomes dominant at lower redshifts, as expected. Across the redshift range studied (\(z = 6\) to \(z = 40\)), the IGM contribution is subdominant in most cases, with the only exception being for \(f_{\text{PBH}} = 10^{-4}\) and \(M_{\text{PBH}} \geq 10 \, \Msun\), where the IGM briefly dominates the emission at higher redshifts. Notably, for higher PBH masses, the redshift of the transition from IGM-dominated to halo-dominated emission shifts to lower values, reflecting the delayed onset of efficient accretion within halos for more massive PBHs. Nevertheless, for such low fractions, the total emission from both the IGM and halos is negligible in comparison to observational constraints.

Interestingly, for certain configurations with \(f_{\text{PBH}} > 10^{-4}\), the IGM contribution increases rather than decreases with decreasing redshift. This behaviour is observed, for example, for \(M_{\text{PBH}} = 1 \, \Msun\) with \(f_{\text{PBH}} \geq 10^{-1}\), \(M_{\text{PBH}} = 10 \, \Msun\) with \(f_{\text{PBH}} \geq 10^{-2}\), and \(M_{\text{PBH}} = 33 \, \Msun\) with \(f_{\text{PBH}} \geq 10^{-2}\). However, even in these cases, the IGM contribution remains negligible compared to the dominant emission from halos.

This counterintuitive increase in IGM emission at lower redshifts can be understood by examining the interplay between accretion dynamics and the evolving density profiles of individual halos. As illustrated in Fig.~\ref{fig:density_profiles}, the central densities of halos decrease with decreasing redshift for a given halo mass. This trend reflects the physical evolution of gas within halos, as the interplay between cooling, accretion, and internal feedback processes shapes the density structure over time. In contrast, the IGM remains less affected by such processes, leading to its relative contribution to X-ray emission becoming more noticeable under specific conditions.

Additionally, this behaviour is closely tied to the balance between the number of PBHs and their individual luminosities. For higher \(f_{\text{PBH}}\), the overall contribution of the IGM is naturally enhanced due to the larger population of PBHs accreting in the diffuse medium. At the same time, for \(M_{\text{PBH}} \geq 10 \, \Msun\), the relative importance of IGM accretion increases further as halo accretion becomes less efficient due to the decreasing gas densities at lower redshifts. This balance is particularly relevant at high fractions \(f_{\text{PBH}}\) because the combined luminosity from a larger number of accreting PBHs in the IGM can partially counteract the suppressed accretion rates within halos.

However, even in these scenarios, the total X-ray emission from the IGM remains significantly lower than that from halos, as the latter dominates the total intensity across all configurations. 

In addition to the soft X-ray background, we also investigated the contribution of PBHs to the hard X-ray background (2 -- 10 keV). The constraints for \( M_{\text{PBH}} = 33 \, \Msun \) and \( 100 \, \Msun \) are identical to those derived from the soft X-ray background, with \( f_{\text{PBH}} \leq 0.001 \). However, for lower masses, the constraints are less restrictive in the hard X-ray regime. Specifically, \( M_{\text{PBH}} = 1 \, \Msun \) is allowed for fractions \( f_{\text{PBH}} \leq 0.1 \), while \( M_{\text{PBH}} = 10 \, \Msun \) is consistent with fractions \( f_{\text{PBH}} \leq 0.01 \). These findings suggest that the hard X-ray background provides complementary but generally weaker constraints compared to the soft X-ray background for low-mass PBHs.

We refine our calculations of \( f_{\text{PBH}} \) by deriving tighter limits that satisfy the observed unresolved soft X-ray background constraints and hence provide more precise results. Using the Makino+1998 and simple isothermal profiles, which yield consistent results, we find that for \( M_{\text{PBH}} = 1 \, M_\odot \), the refined constraint is \( f_{\text{PBH}} \leq 0.007 \), while for \( M_{\text{PBH}} = 10 \, M_\odot \), \( M_{\text{PBH}} = 33 \, M_\odot \), and \( M_{\text{PBH}} = 100 \, M_\odot \), the constraints are \( f_{\text{PBH}} \leq 0.0008 \), \( f_{\text{PBH}} \leq 0.0006 \), and \( f_{\text{PBH}} \leq 0.0007 \), respectively. These refined fractions allow PBHs to explain up to 99, 93, 80, and 91 per cent of the observed excess in the soft X-ray background, while contributing approximately 33, 37, 33, and 39 per cent to the observed excess in the hard X-ray background.

Interestingly, the results reveal that for the same fraction of PBHs (\( f_{\text{PBH}} \)), smaller masses (\( M_{\text{PBH}} \)) can produce slightly higher emissions compared to larger masses. This behaviour arises because smaller PBHs, for a fixed \( f_{\text{PBH}} \), correspond to a larger number density, which enhances the total accretion luminosity. However, this effect is not universally dominant across all PBH fractions but becomes significant only for very small fractions of DM composition (\( f_{\text{PBH}} \ll 0.01 \)), where the increased number of PBHs at lower masses compensates for their reduced individual luminosities. For larger fractions (\( f_{\text{PBH}} \gtrsim 0.01 \)), the luminosity per PBH begins to dominate, leading to higher contributions from more massive PBHs. This highlights the complex interplay between PBH mass, abundance, and accretion efficiency in determining their contribution to cosmic backgrounds.

\begin{figure*}
    \centering
    \includegraphics[width=\textwidth]{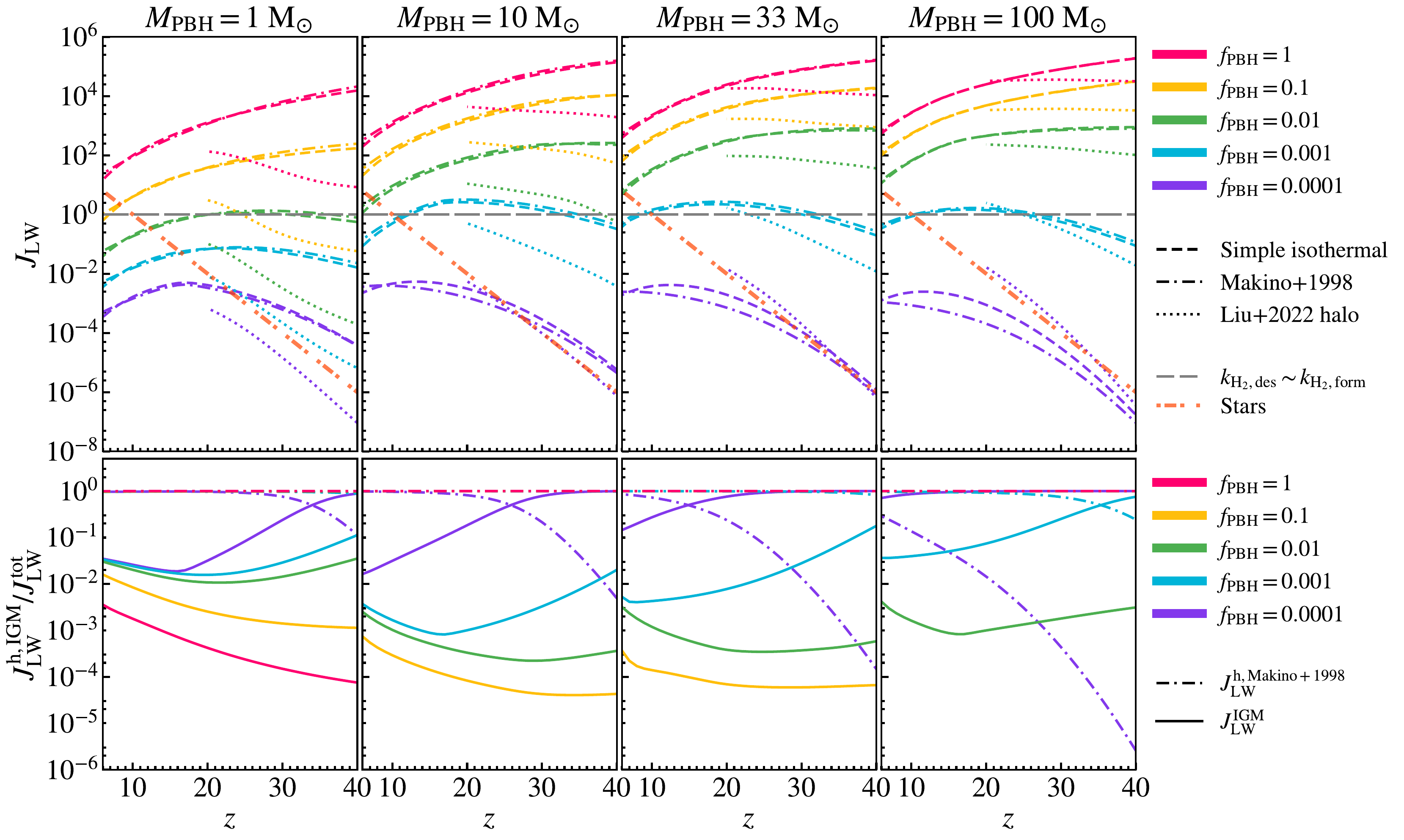}
    \caption{Background intensity of LW radiation (\( J_{\rm{LW}} \)) produced by accreting high-\( z \) PBH sources. The format of panels, colours, and line styles is identical to Fig.~\ref{fig:soft_Xray}: each column corresponds to a different PBH mass, colours represent \( f_{\text{PBH}} \) values, and line styles denote the density profiles considered (simple isothermal, Makino+1998, and Liu+2022). The grey dashed line represents the critical value \( J_{\rm{LW}} \sim 1 \), where the destruction rate of H$_2$ equals the formation rate, while the orange dashed-dotted line shows the contribution from stellar sources for reference.
        The bottom row illustrates the relative contributions of halos (\( J^{\text{h}}_{\text{LW}} \)) and the IGM (\( J^{\text{IGM}}_{\text{LW}} \)) to the total LW background intensity (\( J^{\text{tot}}_{\text{LW}} \)). These are represented as \( J^{\text{h}}_{\text{LW}}/J^{\text{tot}}_{\text{LW}} \) (dashed-dotted lines) and \( J^{\text{IGM}}_{\text{LW}}/J^{\text{tot}}_{\text{LW}} \) (solid lines), respectively. The relative contributions are computed using the same PBH masses and density profile configurations as in the top row. For \( f_{\text{PBH}} > 0.0001 \), the halo contribution dominates at all redshifts, while for \( f_{\text{PBH}} = 0.0001 \), the IGM contribution becomes significant at smaller redshifts, especially for larger PBH masses. }
    \label{fig:LW}
\end{figure*}

\subsubsection{Lyman-Werner Background}\label{subsubsec:lw}

The LWB plays a critical role in the early Universe by dissociating H$_2$, the primary coolant in low-mass halos, thereby regulating star formation and the collapse of structures. The critical threshold \( J_{\rm{LW}} \sim 1 \), where the H$_2$ destruction rate equals its formation rate, marks a tipping point for suppressing small-scale structure formation. PBHs contributing significantly to the LWB could therefore have profound implications for cosmic evolution by delaying or preventing the formation of stars and galaxies in high-redshift environments. A possible example for such efficient LWB delay of early star formation is the massive Pop III candidate GLIMPSE-16043 at $z=6.5$, recently discovered by JWST \citep{Fujimoto_2025}.

Our analysis, shown in Fig. \ref{fig:LW}, quantifies the PBH contribution to \( J_{\rm{LW}} \) for various masses and fractions. For the simple isothermal and Makino profiles, PBHs with \( M_{\text{PBH}} = 1 \, \Msun \) are ruled out for \( f_{\text{PBH}} \geq 10^{-2} \), while those with \( M_{\text{PBH}} = 10 \, \Msun \), \( 33 \, \Msun \), and \( 100 \, \Msun \) are excluded for \( f_{\text{PBH}} \geq 10^{-3} \). Importantly, these constraints are reached at \( z \gtrsim 25 \), before molecular cooling ceases to dominate as the primary cooling mechanism in minihalos. The Liu profile provides less restrictive constraints for \( M_{\text{PBH}} = 1 \, \Msun \) and \( 10 \, \Msun \), ruling out \( f_{\text{PBH}} \geq 10^{-1} \) and \( f_{\text{PBH}} \geq 10^{-2} \), respectively, while remaining consistent with the tighter limits for \( M_{\text{PBH}} = 33 \, \Msun \) and \( 100 \, \Msun \) (\( f_{\text{PBH}} > 10^{-3} \)). This relaxation of constraints for the Liu profile stems from the same characteristics discussed in the context of the X-ray background, where its lower gas densities for minihalos at \( z \gtrsim 20 \) lead to reduced accretion luminosities compared to the other profiles.

The bottom panels of Fig.~\ref{fig:LW} illustrate the relative contributions of halos (\( J^{\text{h}}_{\rm{LW}} \)) and the IGM (\( J^{\text{IGM}}_{\rm{LW}} \)) to the total LWB intensity for various \( f_{\text{PBH}} \) and \( M_{\text{PBH}} \). For \( f_{\text{PBH}} = 10^{-4} \), the IGM contribution dominates the LWB over a range of redshifts \( z < 40 \) across all PBH masses. However, it transitions to being halo-dominated at lower redshifts in all cases except for \( M_{\text{PBH}} = 100 \, \Msun \), where the IGM contribution remains dominant even at \( z = 6 \). In this specific scenario, the IGM exceeds the halo contribution by approximately \( 42\) per cent, accounting for \( 71\) per cent of the total LWB intensity compared to \( 29\) per cent from halos. Despite this, the IGM contribution does not influence the derived constraints, as it remains well below the theoretical LWB limit.

For \( M_{\text{PBH}} = 100 \, \Msun \) with \( f_{\text{PBH}} = 10^{-3} \), there is also a range at \( z < 40 \) where the IGM briefly dominates the total intensity. Nonetheless, by \( z = 6 \), the halo contribution becomes dominant again, as occurs in all other configurations except for \( M_{\text{PBH}} = 100 \, \Msun \) with \( f_{\text{PBH}} = 10^{-4} \).

The relative increase in the IGM contribution with decreasing redshift, observed for specific PBH masses and fractions, follows the same explanation outlined in the X-ray section. It results from the interplay between the evolving density profiles of halos and accretion dynamics. While this effect is notable, the IGM contribution remains negligible compared to that of halos in all cases that are relevant for setting constraints. Additionally, as in the X-ray background case, for higher PBH masses, the transition from IGM dominance to halo dominance occurs at progressively lower redshifts. This effect can be explained by the interplay between PBH mass, their number density for a given \(f_{\text{PBH}}\), and the evolving density structures of the IGM and halos. For more massive PBHs, the number of such objects is lower due to the fixed fraction of DM they represent, which reduces their cumulative emission in the diffuse IGM. Simultaneously, their higher individual accretion rates in halos compensate for their lower abundance, resulting in a stronger contribution from halos at lower redshifts. 

Moreover, while the density profiles of halos evolve and decrease with redshift, the central densities of the most massive halos remain sufficiently high to sustain efficient accretion at lower redshifts. In contrast, the IGM density decreases more rapidly with time, diminishing its relative contribution. This dynamic results in the transition to halo-dominated emission occurring at progressively lower redshifts for higher PBH masses, reflecting the combined effects of hierarchical structure formation, the balance between the number and efficiency of accreting PBHs, and the evolving density profiles of the IGM and halos.

Exceeding the \( J_{\rm{LW}} \sim 1 \) threshold at \( z \gtrsim 25 \) would disrupt molecular cooling in minihalos, critically delaying or suppressing the formation of Population III stars. This makes the constraints on \( f_{\text{PBH}} \) particularly significant, as molecular cooling remains the dominant mechanism for gas condensation in minihalos until \( z \sim 10-15 \), when atomic hydrogen cooling begins to take over in more massive halos (\(M_{\rm{h}} \gtrsim 10^8 \, \Msun\)). Any excess contribution to the LWB beyond the allowed limits at high redshifts would drastically alter the timeline of early star formation, delaying the onset of stellar populations and impacting the formation of early galaxies.

By comparing these findings with the constraints derived from the CXB, we observe that the limits in terms of orders of magnitude remain consistent. However, when considering the more refined constraints, an exception arises: for \( M_{\text{PBH}} = 10 \, \Msun \) and \( f_{\text{PBH}} = 0.0008 \), the LWB intensity exceeds the theoretical limit of \( J_{21} \sim 1 \) at \( z \sim 25 \), reaching \( J_{21} \sim 1.15 \). Consequently, the constraint for this mass is tightened to \( f_{\text{PBH}} \leq 0.0006 \). 

This refinement highlights the complementarity between the CXB and LWB in probing PBH scenarios, as the general consistency between these independent backgrounds strengthens the robustness of the derived constraints while also providing additional precision in specific cases.

Earlier work by \cite{Liu_2022} concluded that the LWB does not impose significant constraints on PBHs. However, we find that this conclusion arises from a coding implementation issue in their modelling, which underestimated the PBH contributions to \( J_{\rm{LW}} \). Our analysis indicates that the LWB provides robust constraints on PBH scenarios, particularly in regulating the thermal and chemical history of the early IGM and the formation of primordial stars.

While we have referred to the upper limits on LW background fluxes as strict constraints, a super-critical LW background (\( J_{\text{LW}} > 1 \)) would, in principle, be possible, shifting the onset of Pop III star formation from minihalos to atomic cooling halos and delaying metal enrichment \citep{Schauer_2021}. However, JWST observations suggest that metal enrichment was already in place for some of the highest-$z$ galaxies \citep{Curti_2022}, making this scenario less likely. In contrast, observational constraints from X-ray and radio backgrounds impose strict upper limits that cannot be exceeded, as they rely on direct measurements. Meanwhile, the LW background constraints are inferred indirectly from their impact on structure formation. Nonetheless, the constraints derived from the X-ray and LW backgrounds are mutually consistent in order of magnitude.

\subsubsection{Radio Background}

\begin{table*}[h!]
\centering

\caption{Brightness temperature and radio excess parameter at \(z = 18\) for different PBH masses and fractions.}
\label{tab:Tb_and_Ar}
\begin{tabular}{cccccccc}
\hline\hline
\(M_{\text{PBH}}\) (\(M_\odot\)) & \(f_{\text{PBH}}\) & \multicolumn{2}{c}{\(T_b\) (K)} & \multicolumn{2}{c}{\(A_r\)} \\
\cmidrule(lr){3-4} \cmidrule(lr){5-6}
 &  & Simple & Makino+1998 & Simple & Makino+1998 \\
\hline
1   & \(0.0001\) & \(1.50 \times 10^{-3}\) & \(1.43 \times 10^{-3}\) & \(2.59 \times 10^{-5}\) & \(2.48 \times 10^{-5}\) \\
    & \(0.007\)  & \(2.80 \times 10^{-1}\) & \(4.64 \times 10^{-2}\) & \(4.84 \times 10^{-3}\) & \(8.02 \times 10^{-4}\) \\
    & \(0.01\)   & \(1.60 \times 10^{-1}\)  & \(1.53 \times 10^{-1}\)  & \(2.77 \times 10^{-3}\)  & \(2.65 \times 10^{-3}\)  \\
    & \(1.0\)    & \(5.44 \times 10^{1}\)   & \(5.15 \times 10^{1}\)   & \(9.40 \times 10^{-1}\)  & \(8.90 \times 10^{-1}\)  \\
\hline
10  & \(0.0001\) & \(1.06 \times 10^{-2}\) & \(1.01 \times 10^{-2}\) & \(1.84 \times 10^{-4}\) & \(1.74 \times 10^{-4}\) \\
    & \(0.0006\) & \(2.62 \times 10^{-1}\) & \(4.18 \times 10^{-2}\) & \(4.53 \times 10^{-3}\) & \(7.23 \times 10^{-4}\) \\
    & \(0.01\)   & \(2.50 \times 10^{0}\)   & \(2.60 \times 10^{0}\)   & \(4.32 \times 10^{-2}\)  & \(4.49 \times 10^{-2}\)  \\
    & \(1.0\)    & \(6.14 \times 10^{2}\)   & \(6.19 \times 10^{2}\)   & \(1.06 \times 10^{1}\)  & \(1.07 \times 10^{1}\)  \\
\hline
33  & \(0.0001\) & \(2.42 \times 10^{-2}\) & \(2.32 \times 10^{-2}\) & \(4.19 \times 10^{-4}\) & \(4.01 \times 10^{-4}\) \\
    & \(0.0006\) & \(6.52 \times 10^{-1}\) & \(1.30 \times 10^{-1}\) & \(1.13 \times 10^{-2}\) & \(2.25 \times 10^{-3}\) \\
    & \(0.01\)   & \(1.22 \times 10^{1}\)   & \(1.29 \times 10^{1}\)   & \(2.12 \times 10^{-1}\)  & \(2.23 \times 10^{-1}\)  \\
    & \(1.0\)    & \(1.72 \times 10^{3}\)   & \(1.76 \times 10^{3}\)   & \(2.97 \times 10^{1}\)  & \(3.05 \times 10^{1}\)  \\
\hline
100 & \(0.0001\) & \(4.51 \times 10^{-2}\) & \(4.39 \times 10^{-2}\) & \(7.80 \times 10^{-4}\) & \(7.59 \times 10^{-4}\) \\
    & \(0.0007\) & \(1.67 \times 10^{0}\)  & \(4.14 \times 10^{-1}\)  & \(2.88 \times 10^{-2}\)  & \(7.16 \times 10^{-3}\)  \\
    & \(0.01\)   & \(3.83 \times 10^{1}\)   & \(4.07 \times 10^{1}\)   & \(6.63 \times 10^{-1}\)  & \(7.03 \times 10^{-1}\)  \\
    & \(1.0\)    & \(4.05 \times 10^{3}\)   & \(4.23 \times 10^{3}\)   & \(7.01 \times 10^{1}\)  & \(7.30 \times 10^{1}\)  \\
\hline
\end{tabular}

\tablefoot{The values shown correspond to the same PBH masses and fractions used in Fig.~\ref{fig:Tbradio}, computed using the simple isothermal and Makino+1998 halo density profiles. Only for PBH masses greater than \(1 \, \rm{M}_{\odot}\) with a fraction \(f_{\mathrm{PBH}} = 1\), the radio excess parameter \(A_r\) satisfies the EDGES constraint (\(1.9 < A_r < 418\); \citealt{Fialkov_2019}), which is consistent with the observed 21-cm absorption signal and a required excess radio background relative to the CMB.}
\end{table*}

The radio background brightness temperature (\(T_b\)) provides crucial insights into the contributions of accreting PBHs to the CRB. Observations from ARCADE 2 reveal a significant excess at low frequencies that cannot be explained by known astrophysical sources \citep{Fixsen_2011, Condon_2012}. Specifically, any Bremsstrahlung contributions from high-$z$ HII regions, powered by star formation in the pre-reionisation universe, would be limited to less than 10 per cent of the signal measured by ARCADE 2 \citep{Liu_2019}.

\begin{figure*}
    \centering
    \includegraphics[width=\textwidth]{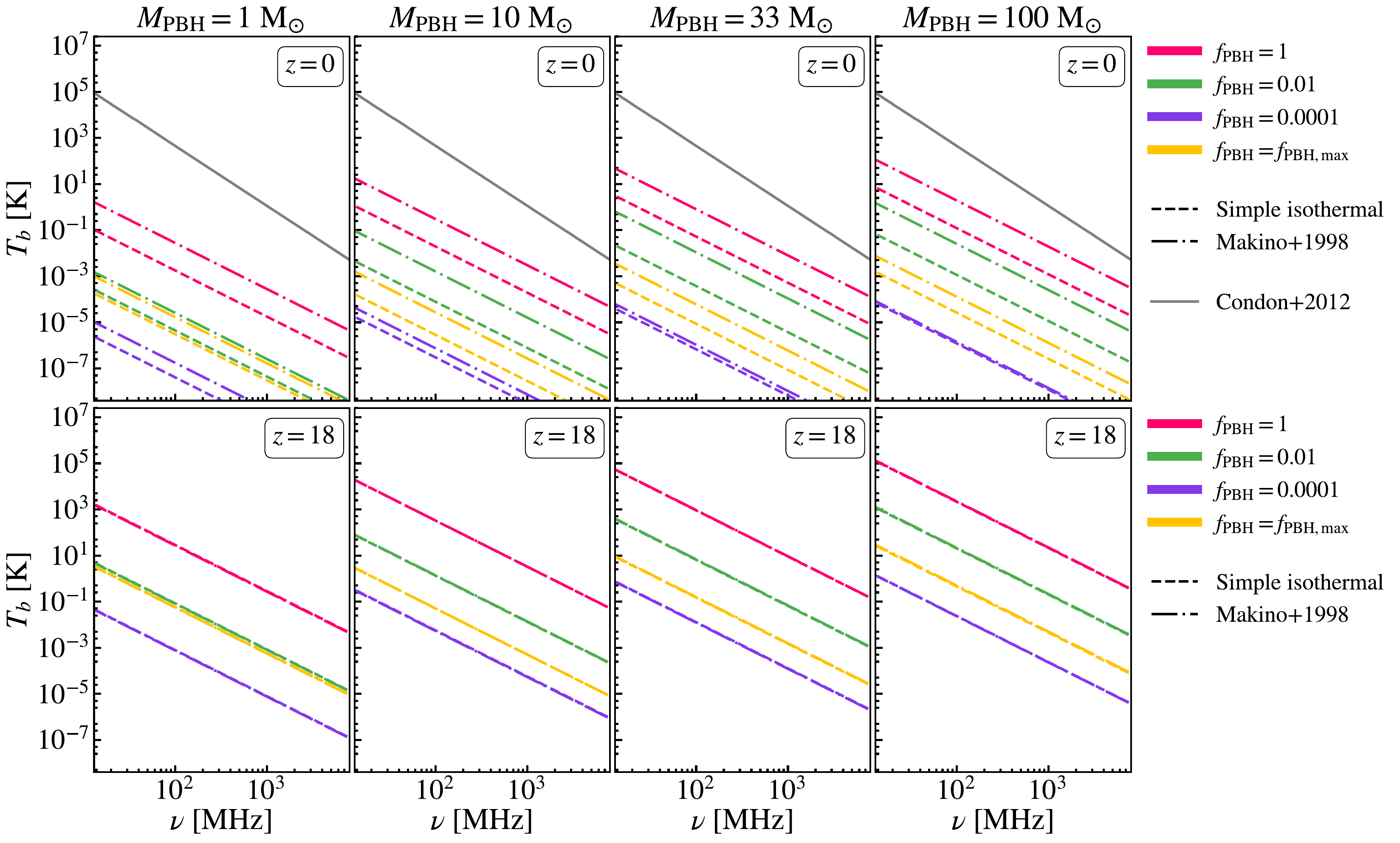}
    \caption{Brightness temperature of the radio background (\(T_b\)) as a function of frequency for PBHs accreting in halos, evaluated at \(z = 0\) (top row) and \(z = 18\) (bottom row). Columns correspond to different PBH masses (\(M_{\text{PBH}} = 1, 10, 33, 100 \, \Msun\)), while the line styles (solid and dashed-dotted) represent the simple isothermal and Makino+1998 density profiles, respectively. The colour scheme indicates \(f_{\text{PBH}}\): red for 1, green for \(10^{-2}\), purple for \(10^{-4}\), and gold for \(f_{\text{max}}\). The maximum fraction of DM allowed for each PBH mass, \(f_{\text{max}}\), is determined from CXB observational and LWB theoretical constraints: \(7 \times 10^{-3}\) for \(M_{\text{PBH}} = 1 \, \Msun\), \(6 \times 10^{-4}\) for \(M_{\text{PBH}} = 10 \, \Msun\) and \(33 \, \Msun\), and \(7 \times 10^{-4}\) for \(M_{\text{PBH}} = 100 \, \Msun\). In the top panels, the grey continuous line corresponds to the fit for the observed radio background excess at \(z = 0\) from \citet{Condon_2012}.}

    \label{fig:Tbradio}
\end{figure*}

At \( z = 0 \), Fig.~\ref{fig:Tbradio} shows the predicted \( T_b \) contributions from PBHs accreting in halos. The results indicate that, even in the extreme case of \( f_{\text{PBH}} = 1 \), the observed excess is not exceeded for both the Makino+1998 and simple isothermal profiles.

The maximum fractions of PBHs allowed by the CXB (observational) and LWB (theoretical) constraints (\(f_{\text{PBH, max}} = 7 \times 10^{-3}\) for \(M_{\text{PBH}} = 1 \, \Msun\), \(6 \times 10^{-4}\) for \(M_{\text{PBH}} = 10 \, \Msun\) and \(33 \, \Msun\), and \(7 \times 10^{-4}\) for \(M_{\text{PBH}} = 100 \, \Msun\)) remain within the observational CRB limits at \(z = 0\). However, their contributions to the CRB excess are negligible, with \(T_b\) at 1420 MHz being at most of the order of \(1\) per cent of the observed excess for the Makino+1998 profile and \(10^{-1}\) per cent for the simple isothermal profile. These results suggest that PBHs, even if they constitute the entirety of DM, cannot significantly contribute to the CRB at the present epoch.

In the second row of Fig.~\ref{fig:Tbradio}, predictions for \(T_b\) at \(z = 18\) are shown. This redshift is of particular interest due to the anomalous 21 cm absorption signal detected by EDGES \citep{Bowman_2018}, corresponding to an observed frequency of \(\nu_{\text{obs}} = 74.8 \, \mathrm{MHz}\). Explaining the depth of this absorption feature requires an enhanced radio background relative to the CMB, quantified by the parameter \(A_r\). This parameter characterises the ratio of the radio background intensity to the CMB intensity at the relevant frequencies and must satisfy \(1.9 < A_r < 418\) \citep{Fialkov_2019}. These constraints are based on the EDGES low-band detection and limits from the Long Wavelength Array 1 (LWA1), which observes the extragalactic radio background at low frequencies \citep{Dowell_2018}.

As seen in Fig.~\ref{fig:Tbradio}, the \(T_b\) values at \(z = 18\) are systematically higher than those at \(z = 0\) across all configurations, reflecting the enhanced accretion rates and denser halo environments at earlier times, where PBHs have a more pronounced impact on the thermal and radiative properties of the gas. Additionally, at \( z = 18 \), the differences between the simple isothermal and Makino+1998 profiles become negligible

Table~\ref{tab:Tb_and_Ar} summarizes the computed \(A_r\) values at \(z = 18\). Configurations that satisfy the EDGES-required range of \(1.9 < A_r < 418\) \citep{Fialkov_2019} correspond only to the extreme cases with \( f_{\text{PBH}} = 1 \) for \( M_{\text{PBH}} > 1 \). Although these configurations appear viable within the CRB limits alone, they are excluded by the tighter CXB and LWB constraints. Furthermore, the maximum PBH fractions permitted by these multi-frequency constraints fall short of producing the required \(A_r\) by 2--4 orders of magnitude. This suggests that, if the EDGES signal is indeed astrophysical in origin, PBHs alone would not be sufficient to account for the enhanced radio background. We also studied the CRB produced by PBHs in the IGM and found that the derived constraints remain unchanged as the contribution within halos is greater.

\cite{Ziparo_2022} also studied the CRB and concluded that PBHs are not significant contributors to the observed radio excess. However, their results differ slightly from ours due to differences in the modelling of radio luminosity (\(L_R\)). While we calculate \(L_R\) based on detailed accretion regimes, including eADAF, standard ADAF, LHAF, and thin disks (as explained in Section \ref{subsec:accretion}), \cite{Ziparo_2022} adopt the fundamental plane relation, which empirically connects radio luminosity, X-ray luminosity, and black hole mass. This approach simplifies the dependency of \(L_R\) on the accretion physics, focusing instead on an observationally derived correlation. Despite these differences, both studies suggest that, even with improved modelling, the contribution of PBHs to the CRB is likely to be small.

These results emphasise the complementarity of multi-frequency analyses in constraining PBH parameter space. While \(T_b\) predictions at \(z = 18\) offer valuable insights into the early contributions of PBHs to radio backgrounds, constraints from CXB and LWB analyses remain the most stringent, ruling out the configurations that could explain the CRB excess.

\begin{figure*}
    \centering
    \includegraphics[width=\textwidth]{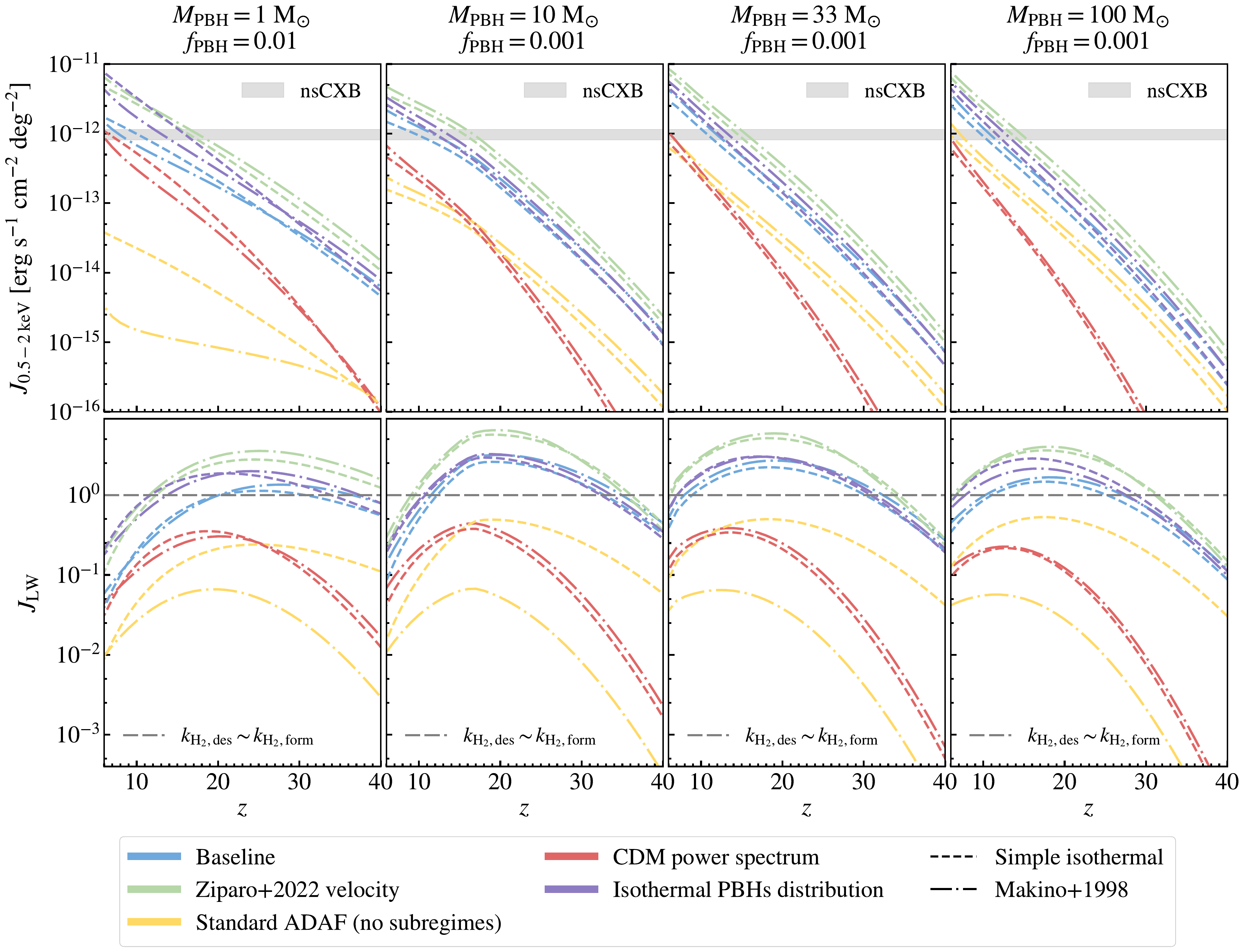}\caption{
Cumulative soft X-ray and Lyman-Werner background intensities from PBHs accreting within halos under different model assumptions. Each column corresponds to a specific PBH mass (\(M_{\text{PBH}} = 1 \, \Msun\), \(10 \, \Msun\), \(33 \, \Msun\), and \(100 \, \Msun\)) and the \(f_{\text{PBH}}\) value derived from the baseline model where the observational (\(\mathrm{nsCXB}\)) or theoretical (\( J_{\rm{LW}} \sim 1 \)) thresholds are exceeded. The top row shows the soft X-ray background intensity (\( J_{0.5-2 \, \mathrm{keV}} \)), while the bottom row displays the LW background intensity (\( J_{\rm{LW}} \)). The coloured lines represent different model variations: baseline model (blue), Ziparo+2022 velocity (green), standard ADAF without subregimes (yellow), CDM power spectrum (red), and isothermal PBH distribution (purple). Dashed and dashed-dotted lines correspond to results using the simple isothermal and Makino+1998 density profiles, respectively. The grey shaded region in the top row indicates the observed excess X-ray background from \citet{Cappelluti_2017}, and the dashed grey line in the bottom row marks the critical \( J_{\rm{LW}} \sim 1 \), where molecular hydrogen dissociation balances its formation.
}\label{fig:variations}
\end{figure*}

\begin{figure*}
    \centering
    \includegraphics[width=\textwidth]{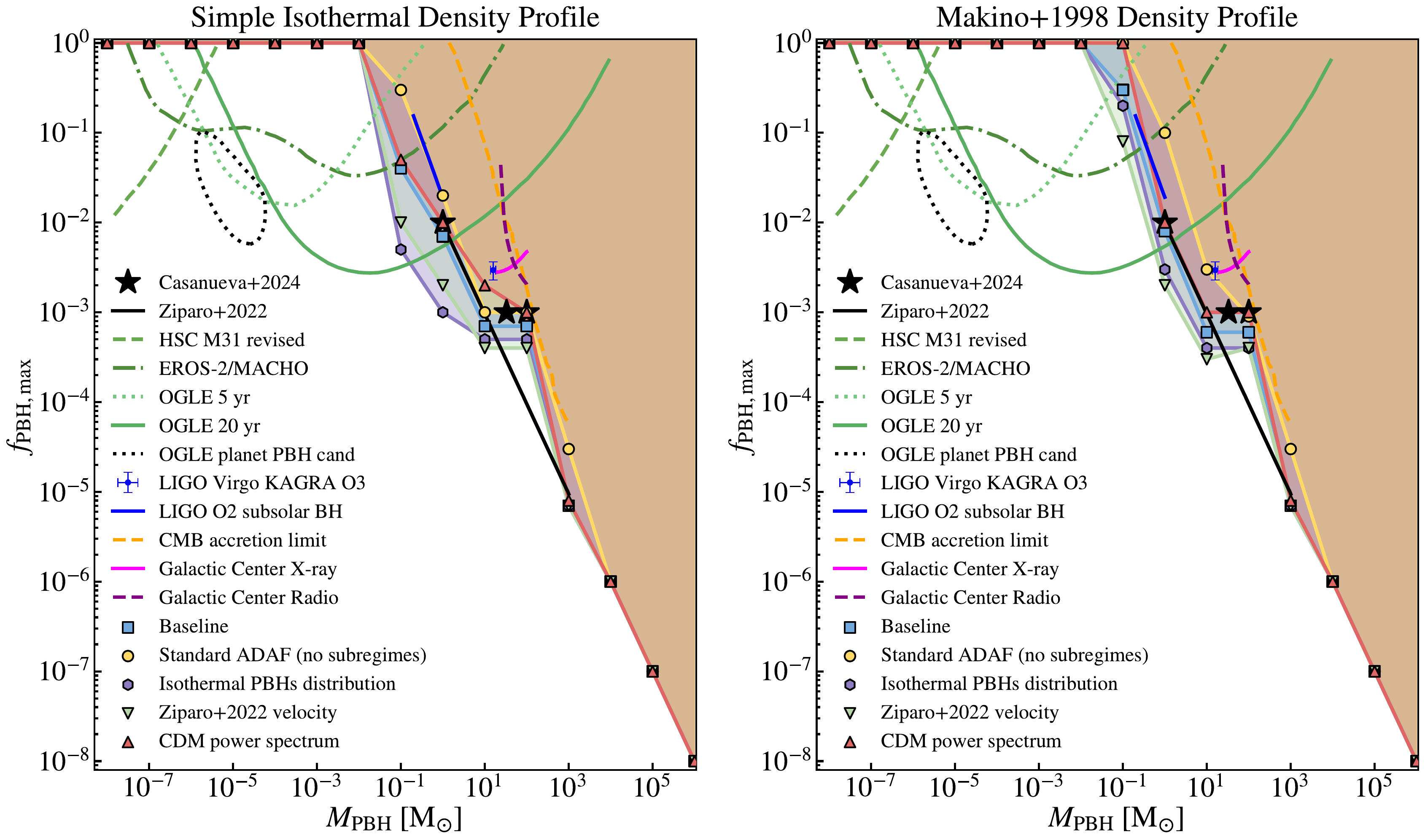}\caption{
Constraints on the maximum allowed PBH DM fraction for different monochromatic PBH masses, including results derived in this work under various modelling assumptions, as well as observational limits and previous theoretical results. We consider two gas density profiles within halos: a simple isothermal (left panel) and the Makino+1998 profile (right panel). For each case, we compute the allowed PBH abundance under different modelling scenarios: baseline (blue squares), standard ADAF without subregimes (yellow circles), isothermal PBH spatial distribution (purple hexagons), Ziparo+2022 velocity prescription (green downward triangles), and CDM power spectrum (red upward triangles). Maximum fractions are computed at one-decade intervals in mass across the range $10^{-8}\,\rm{M}_\odot$ to $10^{6}\,\rm{M}_\odot$. Each coloured marker with a black edge corresponds to the maximum allowed PBH fraction obtained by applying the constraint methodology developed in this work, under a specific model variation. Solid lines connecting the markers and shaded regions, all shown in the same colour as their corresponding markers, are included to aid visual interpretation: although intermediate values were not explicitly computed, the region above each evaluated point can be reliably considered excluded under the corresponding assumptions.
In both panels, our constraints are compared with observational limits, with the results from \citet[solid black line]{Ziparo_2022}, which represent independent theoretical bounds derived from modelling PBH emission contributing to the cosmic X-ray background using a different formalism, and with the results from CV2024 (black stars). While all other constraints correspond to upper bounds on \( f_{\mathrm{PBH}} \), the CV2024 markers indicate fractions for which significant PBH-induced feedback effects are already present in the gas within halos at \( z \sim 23 \). Observational constraints, all derived assuming a monochromatic PBH mass function, include microlensing limits from the Subaru M31 survey \citep{Niikura_2019b} updated by \citet[green dashed line]{Kusenko_2020}, from the Expérience pour la Recherche d’Objets Sombres/MAssive Compact Halo Objects (EROS-2/MACHO) \citep[green dot-dashed line]{Tisserand_2007}, from the OGLE 5 yr survey \citep[dotted green line]{Niikura_2019a}, and from the OGLE 20 yr survey \citep[solid green line]{Mroz_2024}. A dotted black line shows the 95\% confidence region assuming the six ultrashort-timescale OGLE events are due to planetary-mass PBHs \citep{Niikura_2019a}. Additional constraints include the PBH fraction inferred under the assumption that all BH mergers detected during the third observing run (O3) of the LIGO-Virgo-KAGRA Collaboration originate from PBHs \citep[blue point with error bars]{Wong_2021}, subsolar-mass PBH binary merger limits from LIGO O2 \citep[solid blue line]{Abbott_2019}, CMB accretion limits from \citet[orange dashed line]{Poulin_2017}, and Galactic Centre constraints from X-ray (magenta solid) and radio (purple dashed) observations \citep{Manshanden_2019}.
}\label{fig:constraints}
\end{figure*}

\subsection{Effects of model variations on inferred backgrounds and PBH abundances}\label{subsec:model_variations}

To evaluate the influence of different modelling assumptions on the derived constraints for PBHs, we investigate four distinct variations of the baseline model outlined in Section~\ref{sec:baseline_model}. These variations are designed to emphasise the importance of key assumptions and assess their impact on the predicted cosmic backgrounds. The analysis and results presented in Fig.~\ref{fig:variations} show both the soft X-ray and Lyman-Werner background intensities generated by PBHs accreting within halos. Each column corresponds to a specific PBH mass—1, 10, 33, and 100 \(\Msun\)—and the \(f_{\text{PBH}}\) value derived from the baseline model at which the observational threshold, discussed in Section~\ref{subsubsec:xray}, or the theoretical threshold, described in Section~\ref{subsubsec:lw}, is exceeded. The first row shows the cumulative soft X-ray background intensity, while the second row displays the LW background intensity. For all results, accretion in halos is modelled using two gas density profiles: the simple isothermal and Makino+1998 formulations, as indicated in the figure. The third density profile, derived from the rescaled halo in Liu+2022, is not included in this analysis as it does not yield the strongest constraints on \(f_{\text{PBH}}\).

Below, we outline the key features of each variation:

\begin{itemize}
    \item Baseline model: As described in Section~\ref{sec:baseline_model}, this model assumes that PBHs constitute DM distributed according to an NFW density profile, with the HMF modified to account for the presence of PBHs. The characteristic velocity is taken as the virial velocity, which reflects the gravitational potential and velocity dispersion within the halo (blue lines in Fig. \ref{fig:variations}).

    \item Ziparo+2022 velocities: Instead of the virial velocity, the sound speed is adopted, and the relative velocity between PBHs and baryons is set to zero. The sound speed is expressed as:
    \begin{equation}
    c_s = \left(\frac{k_B T_{\text{vir}}}{\mu m_p}\right)^{1/2} \sim 8.3 \, T_{\text{vir,4}}^{1/2} \, \text{km/s},
    \end{equation}

    where \(T_{\text{vir,4}}\) is the virial temperature in units of \(10^4\) K. This assumes hydrodynamical equilibrium, which enhances accretion rates (green lines in Fig. \ref{fig:variations}).

    \item Standard ADAF (no sub-regimes): This variation only considers the standard ADAF and thin disk models, excluding the eADAF and LHAF sub-regimes. This leads to substantially weaker X-ray backgrounds, as the exclusion of LHAF contributions removes a key component of accretion luminosity, which, in this case, plays a dominant role in high-density regions within halos (yellow lines in Fig. \ref{fig:variations}).

    \item CDM power spectrum: The unmodified CDM HMF is adopted, neglecting the effects of PBHs on the mass function. This reduces sensitivity to \(f_{\text{PBH}}\), though the impact on constraints is within an order of magnitude compared to the baseline model (red lines in Fig. \ref{fig:variations}).  

    \item Isothermal PBH distribution: Following \citet{Zhang_2024a}, this variation assumes an isothermal density profile for PBHs instead of the NFW profile used in the baseline model. This results in enhanced accretion rates on smaller scales due to the smoother spatial distribution of PBHs (purple lines in Fig. \ref{fig:variations}).
\end{itemize}

The analysis of model variations highlights several key trends and sensitivities in the derived constraints for \(f_{\text{PBH}}\), emphasising the importance of exploring these variations to understand the dependence of our results on different modelling considerations. When adopting the CDM power spectrum and assuming the Makino+1998 density profile, the constraints become moderately less restrictive across all PBH masses, consistently increasing by a factor of approximately 2 compared to the baseline model while remaining within the same order of magnitude. For instance, for \(M_{\text{PBH}} = 1 \, \Msun\), the combined X-ray and LW constraint shifts from \(f_{\text{PBH}} < 10^{-2}\) to \(f_{\text{PBH}} < 2 \times 10^{-2}\). Similarly, for \(M_{\text{PBH}} = 10 \, \Msun\), \(33 \, \Msun\), and \(100 \, \Msun\), the constraints change from \(f_{\text{PBH}} < 10^{-3}\) to \(f_{\text{PBH}} < 2 \times 10^{-3}\). Importantly, when using the Simple isothermal profile, the constraints derived under the CDM power spectrum are fully consistent with those obtained with the Makino+1998 profile. This uniform relaxation across all masses reflects the relatively limited impact of the unmodified CDM power spectrum on the HMF at lower \(f_{\text{PBH}}\). As discussed in Section~\ref{subsubsec:HMF}, the HMF becomes more significantly modified only at higher PBH fractions (\(f_{\text{PBH}} \sim 1\)), where isocurvature perturbations dominate. These findings suggest that the PBH-induced changes to the HMF are minimal for the parameter space considered, further supporting the robustness of the baseline model's assumptions for the HMF modification.

The ADAF without the sub-regimes model produces the most pronounced deviations, particularly for low-mass PBHs, under the Makino+1998 density profile. For \(M_{\text{PBH}} = 1 \, \Msun\), the X-ray constraint is relaxed by an entire order of magnitude, shifting from \(f_{\text{PBH}} < 10^{-2}\) to \(f_{\text{PBH}} < 3 \times 10^{-1}\). This result implies that the fraction of DM that could consist of PBHs increases from at most \(1\) to \(30\) per cent. For higher masses (\(M_{\text{PBH}} \geq 10 \, \Msun\)), the constraints remain within the same order of magnitude, with the largest variations observed in the LW background. For example, for \(M_{\text{PBH}} = 33 \, \Msun\), the LW constraint shifts from \(f_{\text{PBH}} < 10^{-3}\) to \(f_{\text{PBH}} < 5 \times 10^{-3}\), while for \(M_{\text{PBH}} = 100 \, \Msun\), the LW constraint changes slightly from \(f_{\text{PBH}} < 10^{-3}\) to \(f_{\text{PBH}} < 4 \times 10^{-3}\). Notably, when the Simple isothermal profile is adopted, the relaxation observed for \(M_{\text{PBH}} = 1 \, \Msun\) under the ADAF without the sub-regimes model no longer spans an order of magnitude. Instead, the constraint shifts more modestly from \(f_{\text{PBH}} < 10^{-2}\) to \(f_{\text{PBH}} < 2 \times 10^{-2}\), consistent with the order of magnitude of the baseline model. This reduced impact is attributable to the higher central gas densities in the Simple isothermal profile, which enhance accretion and mitigate the exclusion of the LHAF sub-regime. For higher PBH masses (\(M_{\text{PBH}} \geq 10 \, \Msun\)), the constraints derived with the Simple isothermal profile remain consistent with those obtained using the Makino+1998 profile, showing no significant deviations.

In contrast, the Ziparo+2022 velocity model and the Isothermal PBH distribution yield constraints that are very similar to those of the baseline model, regardless of the chosen density profile. Both models introduce slight tightening of the constraints but do not affect the order of magnitude of the allowed \(f_{\text{PBH}}\). This stability suggests that the relative velocity assumptions and PBH spatial distributions do not significantly influence the inferred constraints, underscoring the robustness of these parameters in shaping the results.

Overall, these findings highlight the importance of modelling assumptions in shaping the constraints on \(f_{\text{PBH}}\), particularly for low-mass PBHs. Notable relaxations arise under the ADAF without the sub-regimes model, especially for \(M_{\text{PBH}} = 1 \, \Msun\) with the Makino+1998 profile, where the exclusion of LHAF significantly weakens the constraints. However, the Simple isothermal profile mitigates these effects due to its higher central gas densities, which enhance accretion. Variations in the CDM power spectrum consistently produce moderate relaxations (factor of 2 across all masses) without changing the order of magnitude, reflecting the limited impact of HMF modifications at low \(f_{\text{PBH}}\). By contrast, variations in velocity assumptions (Ziparo+2022 velocity model) and PBH spatial distributions (Isothermal PBH distribution) show minimal deviations, indicating that these specific assumptions have less influence within the explored parameter space. This analysis validates our results independently of these assumptions and underscores the need to consider such variations to ensure the robustness of the derived constraints.

As stated earlier, all results presented in this work assume a monochromatic PBH mass function. This choice facilitates comparison with previous studies and enables a detailed treatment of accretion processes without requiring additional assumptions about the mass distribution. While such an approximation is well motivated in scenarios where PBHs form from narrow features in the primordial power spectrum, broader mass distributions may weaken constraints by spreading the emission across multiple scales and modifying PBH clustering. Although a full treatment of extended mass functions lies beyond the scope of this paper, it remains an important avenue for future exploration. Constraints on PBH abundance derived under the monochromatic assumption can be reinterpreted for extended mass functions using the formalism of \citet{Bellomo_2018}.

Figure~\ref{fig:constraints} summarises the main results of this work, showing the maximum allowed DM fraction in the form of PBHs, \( f_{\rm PBH,max} \), as a function of monochromatic PBH mass. The figure includes our constraints for different model variations and two halo gas density profiles, along with a comparison to existing observational limits and to the independent theoretical bound from \citet{Ziparo_2022}. Results are shown for a simple isothermal gas density profile (left panel) and for the Makino+1998 profile (right panel). In general, the allowed fraction decreases with all model variations converging to the same \( f_{\rm PBH,max}\) for each individual mass at \( M_{\rm PBH} \geq 10^4\,\mathrm{M}_\odot \), where the IGM contribution dominates due to the decreasing number of haloes massive enough to host at least one PBH. Notably, between \( 33 \, \rm{M}_\odot \) and \( 100 \, \rm{M}_\odot \), \( f_{\rm PBH,max} \) remains roughly constant, which can be attributed to a reduction in the total number of PBHs at these higher masses, compensating for the increased mass of individual PBHs in this specific mass range.

In the isothermal gas profile case, all model variants yield \( f_{\rm PBH,max} = 1 \) for \( M_{\rm PBH} \lesssim 10^{-2}\, \mathrm{M}_\odot \), indicating that our methodology does not constrain PBHs as DM at these scales. However, several observational limits are already more stringent in this mass range. At \( M_{\rm PBH}\sim10^{-1}\; \mathrm{M}_\odot \), the baseline and CDM power spectrum models provide constraints comparable to EROS-2/MACHO, while the standard ADAF model (excluding LHAF and eADAF sub-regimes) yields results consistent with the OGLE 5~yr limit. The models in which PBHs follow an isothermal distribution, or adopt the velocity prescription from \citet{Ziparo_2022}, yield \( f_{\rm PBH,max}(M_{\rm PBH} = 10^{-1}\,\mathrm{M}_\odot) \sim 0.005-0.01 \), below the EROS-2/MACHO and OGLE 5 yr limits.

For the Makino+1998 profile, the constraints are slightly weaker, mainly due to lower central gas densities. Both the standard ADAF (no sub-regimes) and CDM power spectrum variants allow \( f_{\rm PBH} = 1 \) for \( M_{\rm PBH} \lesssim 10^{-1}\, \mathrm{M}_\odot \), while the baseline, isothermal PBH distribution, and Ziparo+2022 velocity models yield \( f_{\rm PBH,max}(M_{\rm PBH} = 10^{-1}\,\mathrm{M}_\odot) \sim 0.1 \), remaining above the OGLE 20~yr and EROS-2/MACHO bounds. At \( M_{\rm PBH} \gtrsim 10^3\,\mathrm{M}_\odot \), all models converge to \( f_{\rm PBH,max} \lesssim 10^{-5} \), independently of the modelling assumptions.

We also compare our results with the theoretical constraint from \citet{Ziparo_2022}, shown as a solid black line. Although the overall shape is similar, the underlying physical assumptions differ substantially. \citet{Ziparo_2022} compute the PBH contribution to the unresolved cosmic X-ray background using a semi-analytic approach based on integrating over the halo population. They adopt the Makino+1998 gas profile and estimate accretion rates for each halo mass bin, assuming a fixed spectral shape for all PBHs: a power law with exponential cutoff at high energies, inspired by ADAF models. The bolometric-to-X-ray conversion is performed using a constant correction factor \( f_X = 0.1 \) in the \(2\text{--}10\,\mathrm{keV}\) band, and no distinction is made between accretion regimes or halo-specific physical conditions. In contrast, our methodology includes regime-dependent spectral modelling, incorporating transitions between ADAF, LHAF, and eADAF, and constructs halo-by-halo spectral energy distributions that depend explicitly on gas density, velocity dispersion, and accretion rate. Despite these differences, both approaches robustly exclude dominant PBH contributions at \( M_{\rm PBH} \gtrsim 10^2\,\mathrm{M}_\odot \), with \( f_{\rm PBH,max} \lesssim 0.001 \), reinforcing the strength of constraints derived from cosmic backgrounds in this mass range. However, our approach produces more detailed constraints and their dependence on different assumptions.

Our results also show good agreement with existing multi-messenger bounds across several mass ranges. At \( M_{\rm PBH} \sim 1\,\mathrm{M}_\odot \), the baseline and standard ADAF-only models in the isothermal gas profile case yield values consistent with the LIGO O2 subsolar merger limit~\citep{Abbott_2019} and the OGLE 20~yr microlensing constraint~\citep{Mroz_2024}, respectively. At \( M_{\rm PBH} \sim 100\,\mathrm{M}_\odot \), our constraints match the CMB accretion bounds from \citet{Poulin_2017}. Among all model variants, the baseline case provides the most restrictive limits across the full mass range, particularly for \( M_{\rm PBH} \sim 1\text{--}10\,\mathrm{M}_\odot \), where it falls below all observational and theoretical bounds considered. This highlights the importance of including sub-regime transitions in the accretion model and accounting for halo-by-halo physical variation when assessing PBH constraints from cosmic backgrounds.

In addition to the limits discussed above, our results can be directly compared with the PBH fractions identified in our previous study, CV2024, shown as black stars in Figure~\ref{fig:constraints}. That analysis focused on the impact of PBHs within individual halos at \( z \sim 23 \), applying the same accretion and emission model used here, which includes all ADAF subregimes (eADAF, ADAF, LHAF) and the thin disk, but implemented in post-processing to halos extracted from one of the simulations in the CIELO suite \citep{CIELO_2025}. Gas densities were modelled using a rescaled high-resolution profile from \citet{Liu_2022}, while characteristic velocities were extracted directly from the simulation particles. Unlike all other constraints shown in the figure, which correspond to maximum allowed values of \(f_{\mathrm{PBH}}\), the CV2024 markers represent the values at which significant feedback effects are already present in the gas. These include temperature increases by a factor of around 300 relative to the no-PBH case, with gas temperatures reaching up to \(1.7\,T_{\mathrm{vir}}\), and a suppression of the central neutral hydrogen abundance to below \(1\%\) of its unperturbed value. Although the two approaches differ in scale and methodology, they yield consistent results. For both the isothermal and Makino+1998 profiles, the baseline models presented in this work yield maximum allowed PBH fractions that lie just below the thresholds identified in CV2024. This agreement across independent frameworks reinforces the robustness of our findings and highlights the complementarity between local feedback signatures and large-scale limits in constraining the PBH abundance.

Finally, we verified that the emission from individual halos remains below current observational detection thresholds and is therefore not removed from the background through source masking. This result holds across all model variations. As a representative example, we considered PBHs with mass $33,\mathrm{M}\odot$, constituting all the DM, embedded in halos of $10^6,\mathrm{M}\odot$ at $z = 30$. This scenario provides a conservative upper bound, as the total emission is dominated by halos with masses below $10^6,\mathrm{M}_\odot$, as discussed in Section~\ref{sec:halo_accretion}. Depending on the model variation, we find total luminosities ranging from approximately $10^{35}$ to $10^{39}$ erg/s in the 0.5–2 keV X-ray band, and from approximately $10^{37}$ to $10^{39}$ erg/s in the 0.044–2.07 eV infrared band. These values remain well below the detection thresholds of current deep field surveys, supporting the validity of comparing our predictions to the unresolved cosmic backgrounds \citep{Cappelluti_2017, Kaminsky_2025}.

\section{Conclusion}\label{sec:conclusion}

Our study places stringent constraints on the maximum fraction of DM that PBHs could constitute while quantifying their contributions to the CXB, LWB, and CRB across different mass ranges. A detailed summary is provided below for reference PBH masses:

\begin{itemize}
    \item \(1 \, \Msun\): If PBHs constituted \(7 \times 10^{-3}\) of the DM, they would fully explain the CXB background (\(99\) per cent). A larger fraction would exceed the observed values. On the other hand, such an abundance of \(1 \, \Msun\) PBHs only explains \(33\) per cent of the hard CXB. Their contribution to the LWB is significant but remains below the critical threshold (\(J_{\rm LW} \sim 1\)) needed to suppress star formation. Contributions to the CRB are negligible.

    \item \(10 \, \Msun\): PBHs are constrained to a maximum DM fraction of \(6 \times 10^{-4}\). In this case, they can contribute up to \(93\) per cent of the soft CXB and \(37\) per cent of the hard CXB. Similarly, their LWB contribution is non-negligible but remains consistent with the \(J_{\rm LW} \sim 1\) threshold, ensuring no significant disruption of molecular cooling. Their impact on the CRB remains insignificant.

    \item \(33 \, \Msun\): PBHs can comprise up to \(6 \times 10^{-4}\) of DM. They would contribute in this case up to \(80\) per cent of the soft CXB and \(33\) per cent of the hard CXB. As with lower masses, their LWB contributions are limited by the \(J_{\rm LW} \sim 1\) threshold, and their CRB contributions are negligible.

    \item \(100 \, \Msun\): PBHs are limited to a maximum DM fraction of \(7 \times 10^{-4}\). These can explain up to \(91\) per cent of the observed soft CXB excess and \(39\) per cent of the hard CXB. Their LWB contributions are significant but adhere to the \(J_{\rm LW} \sim 1\) limit, while their CRB contributions remain negligible.
\end{itemize}

Our study emphasises the importance of detailed modelling and the exploration of variations in key assumptions. We adopt an accretion model that includes distinct subregimes—eADAF, standard ADAF, LHAF, and thin disks—allowing us to capture the diversity of PBH accretion scenarios based on local halo and IGM conditions. Notably, including these subregimes can relax the constraints on low-mass PBHs (\(1 \, \Msun\)) by up to an order of magnitude compared to models without them, due to the enhanced accretion luminosities captured by this approach. We also test the robustness of our results against variations in gas density profiles, velocity prescriptions, emission models, and the PBH impact on the halo mass function.

Building on this framework, we extended our analysis across the full mass range \(10^{-8} \text{--} 10^6\,\rm{M}_\odot\), enabling a direct comparison with a wide array of observational and theoretical constraints. In particular, we find that our baseline model yields constraints more restrictive than all existing bounds in the intermediate mass range \( M_{\rm PBH} \sim 1\text{--}10\,\rm{M}_\odot \), for both density profiles considered. These are slightly stronger than those of \citet{Ziparo_2022}, though within the same order of magnitude. These results also demonstrate that assumptions regarding accretion subregimes, PBH spatial distributions, or velocity prescriptions can shift the allowed \( f_{\rm PBH} \) values by factors of a few, with the most extreme case relaxing constraints by up to an order of magnitude. Altogether, these findings reinforce the value of a comprehensive and physically grounded modelling approach when evaluating the viability of PBHs as dark matter candidates.

While our results align with many previous studies, the combination of detailed accretion modelling, multi-frequency analysis, and systematic exploration of model variations provides a more nuanced understanding of PBHs as DM candidates. This approach refines the allowed parameter space for PBHs and emphasises the role of robust modelling in interpreting their astrophysical implications.

Our constraints, derived from their contributions to multiple cosmic radiation backgrounds, indicate that PBHs cannot constitute the totality of DM. However, their potential to uniquely impact early cosmic history further motivates their in-depth investigation, as their astrophysical signatures could provide key insights into both their nature and the broader evolution of the universe.

\begin{acknowledgements}

NP acknowledges support
from Préstamo BID PICT Raices 2023 Nº 0002. PBT acknowledges partial funding by Fondecyt-ANID 1240465/2024 and Núcleo Milenio ERIS NCN2021\_017. BL gratefully acknowledges funding from the Deutsche
Forschungsgemeinschaft (DFG, German Research Foun-
dation) under Germany’s Excellence Strategy EXC
2181/1 - 390900948 (the Heidelberg STRUCTURES Ex-
cellence Cluster). This project has received funding from the European Union Horizon 2020 Research and Innovation Programme under Marie Skłodowska-Curie Actions (MSCA) grant agreement No. 101086388-LACEGAL. We acknowledge partial support by ANID BASAL project FB210003. We also thank Prof. Dr. Günther Hasinger for his valuable comments and suggestions, which have significantly improved this manuscript.

\end{acknowledgements}

\bibliographystyle{aa} 
\bibliography{aa54032-25}

\newpage
\appendix
\section{Derivation of radiation background equations}
\label{appendix:derivation}

In this appendix, we present the derivation of the equations used to calculate the radiation background intensity arising from PBHs accreting within halos and the IGM, considering the cumulative contribution of sources distributed among cosmic structures.

The physical (proper) flux \( \delta F \), defined as the energy received per unit time and per unit area by an observer at redshift \( z \), is given by:

\begin{equation}
    F = \frac{\delta E}{\delta A \, \delta t},
\end{equation}
where \( \delta E \) represents the energy of a photon packet passing through an area \( \delta A \) over a time interval \( \delta t \) in the observer frame at redshift \( z \). In the case of a diffuse source, the energy and time intervals in the source frame are modified due to cosmological redshift as follows:

\begin{equation}
    \delta E' = \frac{1+z'}{1+z} \delta E, \quad \delta t' = \frac{1+z}{1+z'} \delta t.
\end{equation}

Thus, the observed flux can be rewritten as:

\begin{equation}
    F = \left( \frac{1+z}{1+z'} \right)^2 \frac{1}{\delta A} \frac{\delta E'}{\delta t'}.
    \label{eq:flux_def}
\end{equation}

For a diffuse radiation field, the luminosity can be expressed in terms of the source frame physical emissivity \( \epsilon(z') \), which quantifies the energy emitted per unit volume and per unit time. Considering that the radiation originates from a diffuse medium occupying a physical volume of \( \delta V' = \delta A' \delta \ell' \) at redshift \( z' \), where \( \delta \ell' \) is the line element and \( \delta A' \) represents the cross-sectional area of the cylinder spanned by the light rays connecting the source to the observer, the luminosity takes the form:

\begin{equation}
    \frac{\delta E'}{\delta t'} = \epsilon(z') \delta V' = \epsilon(z') \delta A' \delta \ell'.
\end{equation}

Considering the cosmic expansion, the relationship between the observed cross-sectional area and the comoving area is given by:

\begin{equation}
    \delta A' = \left( \frac{1+z}{1+z'} \right)^2 \delta A.
\end{equation}

Substituting these relations into Eq.~\eqref{eq:flux_def}, the flux for a diffuse source can be expressed as:

\begin{equation}
    F = \left( \frac{1+z}{1+z'} \right)^4 \epsilon(z') \delta \ell'.
    \label{eq:flux_diffuse}
\end{equation}

The total flux, obtained by integrating the contributions from sources along the line of sight up to a maximum redshift \( z_{\max} \), is given by:

\begin{equation}
\begin{aligned}
    F &= \int_{z}^{z_{\max}} \left( \frac{1+z}{1+z'} \right)^4 \epsilon(z') \frac{d\ell'}{dz'} dz' \\
      &= - \int_{z}^{z_{\max}} \left( \frac{1+z}{1+z'} \right)^4 \epsilon(z') c \frac{dt_{\rm U}}{dz'} dz'
      \label{eq:F_dtudz}
\end{aligned}
\end{equation}
where the physical line element \( d\ell' \) is related to the cosmic age \( t_{\mathrm{U}} \) through the relation \( d\ell'/dz' = -c \, dt_{\mathrm{U}}/dz' \).

To account for frequency dependence, the specific flux (energy per unit time, per unit area, per unit frequency) in the observer frame is defined as:

\begin{equation}
    F_{\nu} \equiv  \frac{dF}{d\nu},
    \label{eq:specific_flux}
\end{equation}
where the observed frequency \( \nu \) and the source-frame frequency \( \nu' \) are related by:

\begin{equation}
    \nu = \frac{1+z}{1+z'} \nu'.
\end{equation}

The specific emissivity of the diffuse source is then defined as:

\begin{equation}
    \epsilon_{\nu'} = \frac{d\epsilon}{d\nu'}.
\end{equation}

Substituting these definitions, the total frequency-dependent flux from diffuse sources is expressed as:

\begin{equation}
\begin{aligned}
F_{\nu} &= - \int_{z}^{z_{\text{max}}} \left( \frac{1+z}{1+z'} \right)^{4} 
\frac{d\epsilon(z')}{d\nu'} \frac{d\nu'}{d\nu} c \frac{dt_{U}}{dz'} \, dz' \\
&= - \int_{z}^{z_{\text{max}}} \left( \frac{1+z}{1+z'} \right)^{3} 
\epsilon_{\nu'}(z') \frac{c \, dt_{U}}{dz'} \, dz'.
\end{aligned}
\label{eq:F_nudtdz}
\end{equation}

In addition to cosmic age \( t_{\rm U} \), the flux equations can be reformulated in terms of the comoving distance \( r \), defined from \( z = 0 \) to \( z' \). Given the relation \( dr/dz' = -(1 + z') c \, dt_{\rm U}/dz' \), the flux can be rewritten as:

\begin{equation}
    F = \int_{z}^{z_{\max}} \frac{(1 + z)^4}{(1 + z')^5} \, \epsilon(z') \frac{dr}{dz'} \, dz' \, ,
    \label{eq:F_drdz}
\end{equation}

\begin{equation}
    F_{\nu} = \int_{z}^{z_{\max}} \frac{(1 + z)^3}{(1 + z')^4} \, \epsilon_{\nu'}(z') \frac{dr}{dz'} \, dz' \, .
\end{equation}

Finally, the radiation intensity \( J \), defined as the energy per unit time, per unit area, and per unit solid angle, is related to the flux by the expression \( J = F / (4\pi) \). When considering frequency dependence, an additional dependence per unit frequency is introduced, leading to the definition of the specific intensity \( J_{\nu} \), which is related to the specific flux by \( J_{\nu} = F_{\nu} / (4\pi) \).

To derive Eq.~\eqref{eq:JX}, we begin with Eq.~\eqref{eq:F_drdz} and use the relation \( \epsilon(z') = \epsilon_{\rm cm}(z') (1+z')^3 \), where the comoving integrated emissivity $\epsilon_{\rm cm}(z')$ over a frequency range \([\nu_1, \nu_2]\) is given by:

\begin{equation}
    \epsilon_{\rm cm,[\nu_1,\nu_2]}(z,z') = \int_{\nu_1(z,z')}^{\nu_2(z,z')} \frac{\epsilon_{\nu'}(z')}{(1+z')^3} d\nu' = \int_{\nu_1(z,z')}^{\nu_2(z,z')} \dot{\rho}_{\nu'}(z') d\nu'.
\end{equation}

Substituting this definition into Eq.~\eqref{eq:F_drdz} yields:

\begin{equation}
    J = \frac{1}{4\pi} \int_{z}^{z_{\max}} \frac{(1 + z)^4}{(1 + z')^2} \, \epsilon_{\rm cm}(z') \frac{dr}{dz'} \, dz'.
\end{equation}

Evaluating the expression at \( z=0 \) results in the cancellation of the factor \( (1+z)^4 \). By discretising the integral, we obtain Eq.~\eqref{eq:JX}.

To derive Eq.~\eqref{eq:JLW}, we start from Eq.~\eqref{eq:F_dtudz} and apply the relation \( \epsilon(z') = \epsilon_{\rm cm}(z') (1+z')^3 \). Substituting the cosmic time differential \( dt_{\rm U}/dz' \) with:

\begin{equation}
    \frac{dt_{\rm U}}{dz'} = \frac{dt_{\rm U}}{da'} \left(-\frac{1}{(1+z')^2}\right),
    \label{eq:dt_dz}
\end{equation}
we arrive at the following expression for the background intensity:

\begin{equation}
    J = \frac{1}{4\pi} \int_{z}^{z_{\max}} \frac{(1+z)^4}{(1+z')^3} 
   \epsilon_{\rm cm}(z') c\left. \frac{dt_{\rm U}}{da'} \right|_{z'}
dz'.
\end{equation}

Similarly, to derive Eq.~\eqref{eq:Jnu}, we start from Eq.~\eqref{eq:F_nudtdz} and incorporate the comoving specific emissivity \( \dot{\rho}_{\nu'}(z') \) along with Eq.~\eqref{eq:dt_dz}. Substituting these expressions results in:

\begin{equation}
    J_{\nu} = \frac{1}{4\pi} \int_{z}^{z_{\max}} \frac{(1+z)^3}{(1+z')^2} 
    \dot{\rho}_{\nu'}(z') c\left. \frac{dt_{\rm U}}{da'} \right|_{z'}
dz'.
\end{equation}

\end{document}